\documentclass[twocolumn,footinbib,floatfix,aps,10pt,pra,superscriptaddress]{revtex4-1}
\usepackage[usenames,dvipsnames]{xcolor} 
\usepackage{amsmath,amssymb}
\usepackage{amsfonts}
\usepackage{bm}
\usepackage{bbm}
\usepackage{bbold}
\usepackage{color}
\usepackage[english]{babel}
\usepackage{psfrag,graphicx}
\usepackage{subfigure}
\usepackage{natbib}
\usepackage{soul}
\usepackage{appendix}
\usepackage{enumitem}
\usepackage[applemac]{inputenc} 
\usepackage[normalem]{ulem}
\usepackage{amsthm}

\definecolor{mygrey}{gray}{0.35}
\definecolor{myblue}{rgb}{0.2,0.2,0.8}
\definecolor{myzard}{cmyk}{0,0,0.05,0}
\definecolor{mywhite}{rgb}{1,1,1}
\definecolor{myred}{rgb}{0.9,0.1,0.}
\usepackage[colorlinks=true,citecolor=myblue,linkcolor=myblue,urlcolor=myblue]{hyperref}

\theoremstyle{definition}

\newcommand{\ket}[1]{\vert #1 \rangle} 
\newcommand{\bra}[1]{\langle #1 \vert} 
\newcommand{\ketbra}[2]{\vert #1 \rangle \hspace{-2pt}\langle #2 \vert}

\newcommand{\eref}[1]{(\ref{#1})}
\newcommand{\fref}[1]{figure \ref{#1}}
\newcommand{\Fref}[1]{Fig.\ref{#1}}

\newcommand{\llrr}[1]{\ensuremath{\left( #1\right)}}
\newcommand{\llrrq}[1]{\ensuremath{\left[ #1\right]}}

\newcommand{\bs}[1]{\ensuremath{\boldsymbol{#1}}}

\DeclareMathOperator{\Tr}{Tr}

\begin{document}


\title{Quantum probing beyond pure dephasing}

\author{Dario Tamascelli}
\affiliation{Quantum Technology Lab, Dipartimento di Fisica ``Aldo Pontremoli'', Universit\`a degli Studi di Milano, I-20133 Milano, Italy}
\affiliation{Institut f\"{u}r Theoretische Physik, Albert-Einstein-Allee 11,
Universit\"{a}t Ulm, 89069 Ulm, Germany}
\author{Claudia Benedetti}
\affiliation{Quantum Technology Lab, Dipartimento di Fisica ``Aldo Pontremoli'', Universit\`a degli Studi di Milano, I-20133 Milano, Italy}
\author{Heinz-Peter Breuer}
\affiliation{Physikalisches Institut, Albert-Ludwigs-Universit\"at Freiburg, Hermann-Herder-Str. 3, 79104 Freiburg, Germany}
\affiliation{Freiburg Institute for Advanced Studies (FRIAS), 
Albert-Ludwigs-Universit\"at Freiburg, Albertstr. 19, 79104 Freiburg, Germany}
\author{Matteo G. A. Paris}
\affiliation{Quantum Technology Lab, Dipartimento di Fisica ``Aldo Pontremoli'', Universit\`a degli Studi di Milano,  I-20133 Milano, Italy}
\affiliation{INFN - Sezione di Milano, I-20133 Milano, Italy}

\date{\today}

\begin{abstract}
Quantum probing is the art of exploiting simple quantum systems interacting with a complex environment to extract precise information about some environmental parameters, e.g. the temperature of the environment or its spectral density. Here we analyze 
the performance of a single-qubit probe in characterizing Ohmic bosonic environments at thermal equilibrium. In particular, we analyze the effects of tuning the 
interaction Hamiltonian between the probe and the environment, going beyond the traditional paradigm of pure dephasing. In the weak-coupling and short-time 
regime, we address the dynamics of the probe analytically, whereas numerical 
simulations are employed in the strong coupling and long-time regime. We then
evaluate the quantum Fisher information for the estimation of the cutoff 
frequency and the temperature of the environment. Our results provide clear 
evidence that pure dephasing is not optimal, unless we focus attention to 
short times. In particular, we found several working regimes where 
the presence of a transverse interaction improves the maximum attainable precision, i.e. it increases the quantum Fisher information. We also explore the role of the initial state of the probe and of the probe characteristic frequency in determining  the estimation precision, thus providing quantitative guidelines to design optimized detection to characterize bosonic environments at the quantum level.
\end{abstract}

\maketitle

\section{Introduction}
Being able to characterize the properties of a  complex environment through a simple, small and controllable quantum system  is the leading scope of quantum probing \cite{elliott16,viola16,benedetti14, benedettip14,tama16, cosco17, cappellaro17, petit18}. 
This topic has a natural connection with the theory 
of quantum estimation, where the aim is to be able 
to precisely infer the value of unknown parameters 
through repeated measurements on the system of interest \cite{paris09,braun13,genoni13,toth14,troiani18, seveso19}. 
Indeed, the quality of a quantum probe can be evaluated through the error committed in characterizing parameters of the environment.  The quantum Fisher information (QFI) is a measure 
of this error through the quantum Cram\'er-Rao bound. 
In order to extract the maximum information from a probing scheme,
one needs to optimize the procedure over the preparation of the probe and over the kind of probe-environment interaction. 
We illustrate this problem by focusing on the estimation of the 
cutoff frequency and of the temperature of a bosonic bath with
 an Ohmic-like spectral density by using a single qubit as a quantum probe.
This problem has already been addressed in Ref. \cite{benedetti18} for the specific case of spin-boson model which induces a dephasing
on the qubit dynamics \cite{breuer02}, where it was shown that the optimal initial states of the probe are always the maximally coherent states (in the computational basis), e.g. the eigenstates of the $\sigma_x$ Pauli matrix.
\par
In this work we address the problem whether dephasing is the optimal
interaction for the estimation of environmental parameters. 
As a matter of fact, in a pure dephasing dynamics only the coherences 
of the system can be affected by the interaction with the  environment.  
Other interactions, by allowing all the components of the reduced density matrix of the probe qubit to change,  may lead to a larger gain of information on the environmental features, 
and thus to a more precise estimation of the 
inferred parameter(s). We show that this is indeed the case by considering the QFI related to the estimation of the cutoff frequency of the  spectral density and the environmental temperature.
 In order to shed light on the role of the kind of system-environment interaction and of the probe's initial preparation on the ultimate estimation precision attainable, we analyze 
the behavior of the QFI as a function of time
for different types of probe-bath interactions and different
initial states of the probe. 
To determine the numerically exact evolution of the probe density matrix, we exploit the TEDOPA (Time Evolving Density operator with Orthogonal PolynomiAls) algorithm \cite{tamascelli19,prior10, woods14, woods2015}, which allows for the efficient simulation of spin-boson models. While an exact analytic treatment is possible only for the specific case of pure dephasing dynamics,  perturbative expansions, such as the Time Convolution-Less (TCL) master equation \cite{shibata77,breuer01}, are accurate only in the weak-coupling regime. Moreover, since in our setting we are interested in properties of the environment and not of the system, the general results derived in \cite{Haase18} are not applicable.
\par
Our results show that while dephasing {enhances the estimation precision at very short times, it is never  optimal  at longer times.}
 The optimal initial state of the probe depends on the specific interaction chosen. We moreover bring evidence of the fact that the frequency of the probe qubit has a major impact on the ultimate estimation precision of environmental parameters. The paper is organized as follows. In Section I we introduce the spin-boson model and the spectral density. In Section II we define the quantum Fisher information. Section IV is devoted to the derivation of the QFI in the weak-coupling limit. In Section IV we consider the arbitrary coupling case, and determine an approximate short-time evolution of the QFI in this scenario. The behavior of the QFI over longer times, obtained by numerical t-DMRG techniques, are discussed in Section IV, before  drawing our conclusion and offering perspectives.
\section{The system}
We consider a two-level system (TLS) interacting with a structured bosonic environment. For each environmental mode at frequency $\omega \geq 0$ the annihilation and creation operators $a_\omega, a_\omega^\dagger$ satisfy the commutation relations $[a_\omega,a_{\omega'}^\dagger]=\delta_{\omega,\omega'}$, $[a_\omega,a_{\omega'}]=[a_\omega^\dagger,a_{\omega'}^\dagger]=0$, $\forall \omega,\omega' \geq 0$.  The overall
(system+environment) Hamiltonian is 
\begin{align}
    H_{SE}(\theta) &= {H_S +H_E+H_I(\theta)}, \label{eq:completeHam}\\
    H_S &= \frac{1}{2} \omega_S \sigma_z,\\
    H_E &= \int_0^{\infty} \!\! d\omega \,\omega\, a_\omega^\dagger a_\omega,\\
    H_I(\theta) &= A_S(\theta) \otimes G_E 
\end{align} 
and the
operators 
\begin{align}
A_S(\theta) &=\frac{\sigma_x}{2} \cos\theta  + \frac{\sigma_z}{2} \sin\theta
\label{eq:interaction} \\
G_E &= \int_0^\infty \!\!\! d\omega \, \sqrt{J(\omega)} (a_\omega +a_\omega^\dagger)
\end{align} 
model the  system-environment interaction. Here and in what follows $\sigma_{x,y,z}$ are the Pauli matrices.  When $\theta=\pi/2$,  the dephasing model  is recovered
while for other values of $\theta$ the transverse  (w.r.t. the system free Hamiltonian $H_S$) components come into play such that $H_S$ and $A_S$ no longer
commute, leading  to more involved dynamics for the probe qubit. 

The function $J(\omega): \mathbb{R}^+ \to \mathbb{R}^+$ is defined by the product of the interaction strength between  the system and the environmental mode at frequency $\omega$ and the mode density around $\omega$, and is usually referred to as the \emph{spectral density} (SD). At time $t=0$, system and environment are assumed to be in a factorized state $\rho_{SE}(0) = \rho_S(0) \otimes \rho_E(0)$, where $\rho_S(0)$ is an arbitrary state of the probe, $\rho_E(0) = \otimes_\omega \exp\llrr{- \beta \omega a_\omega^\dagger a_\omega}/\mathcal{Z}_\omega$ is the thermal state of the environment at inverse temperature $\beta = 1/T$, and $Z_\omega$ is the partition function of the mode at frequency $\omega$. Under these assumptions, the spectral density $J(\omega)$ entirely determines the open-system state $\rho_S(t) = \Tr_E\llrrq{\rho_{SE}(t)}$, since it determines 
the two-time correlation function (TTCF)\\
\begin{align}
C(t) &=  \langle G_E(t) G_E(0) \rangle_{\rho_E(0)}  =\langle e^{i H_E t} G_E e^{-i H_E t}  G_E \rangle_{\rho_E(0)}  \nonumber \\
 &= \int_0^{+\infty} \hspace{-3pt}d\omega J(\omega) \llrrq{n_\beta(\omega) e^{i \omega t} +(1+n_\beta(\omega)) e^{-i \omega t}} \nonumber  \\
& = \int_{-\infty}^{+\infty} d\omega e^{i \omega t}j_\beta(\omega).
\end{align}
where $n_\beta(\omega) = 1/(e^{\beta \omega}-1)$ and
\begin{align}
j_\beta(\omega) =  \frac{1}{2} \llrrq{1+\coth\llrr{ \frac{\beta \omega}{2}}}
&\llrrq{J(\omega)\Theta(\omega)-J(-\omega) \Theta(-\omega)}
\end{align}
is a non-negative function that we will refer to as to the \emph{thermalized spectral density} \cite{tamascelli19}. Since the environment is initially in a (Gaussian) thermal state, multi-time correlations are all functions of the TTCF $C(t)$ alone.

In this work we will consider Ohmic
spectral densities of the form
\begin{equation}
    J(\omega) = \frac{\lambda}{\omega_c^{s-1}} \omega^s e^{-\frac{\omega}{\omega_c}},
    \label{eq:ohmic}
\end{equation}
where $\lambda$ is an overall constant, $s>0$ is the Ohmicity parameter, $\omega_c$ indicates the bath cutoff frequency, and we assumed an exponential form of the cutoff. The corresponding TTCF reads
\begin{align}
&C_{\lambda,s,\omega_c,\beta}(t) = \frac{\lambda s! \omega_c^2 }{(1+i \omega_c t)^{s+1}} + \lambda \omega_c^2  \llrr{-\frac{1}{\beta \omega_c }}^{s+1} \times \nonumber \\
&  \llrrq{\Phi^{(s)}\llrr{1+\frac{1+i \omega_c t}{\beta \omega_c}}+\Phi^{(s)}\llrr{1+\frac{1-i \omega_c t}{\beta \omega_c}}},
\end{align}
$\Phi^{(s)}(z)$ being the polygamma function of order $s$.

{In the following we will assume $\omega_c = 1$ and express time and frequency in dimensionless $\omega_c$-based units. We also use natural units $\hbar=k_B=c=1$ throughout the paper.} 
\section{Quantum Fisher information}
Consider a family of quantum states $\{\rho_{\eta}\}$ depending on the parameter $\eta$ which we want to estimate.
The ultimate precision of any unbiased estimator {$\widehat{\eta}$} of the parameter $\eta$ is given by the single-shot
quantum Cram\`er-Rao inequality:
\begin{equation}
\sigma^2[\widehat{\eta}]\geq \frac{1}{Q(\eta)},
 \label{qcrb}
\end{equation}
where  $\sigma^2$ is the variance of the estimator and $Q(\eta)$ is the quantum Fisher information defined as:
\begin{equation}
Q(\eta)=\Tr[\rho_{\eta}L_{\eta}^2].
\end{equation}
 $L_{\eta}$ is the symmetric logarithmic derivative implicitly defined by  
 $\frac{\partial\rho_{\eta}}{\partial\eta}=\frac12 \{L_{\eta},\rho_{\eta}\}$ and $\{\cdot\}$ denotes the anticommutator.
 The QFI thus quantifies  the ability to estimate an unknown parameter by posing a lower bound to the variance of the estimator $\widehat{\eta}$.
 The problem to accurately infer the value of an unknown parameter is strictly connected to the ability 
 to discriminate between states $\rho_{\eta}$ and $\rho_{\eta+ \delta\eta}$, where $\delta\eta$ is an infinitesimal small deviation.  The larger the QFI, 
 he higher is the ability to distinguish between neighboring states (in $\eta$), and the smaller is the error associated to the estimation procedure.  Not surprisingly, thus, $Q(\eta)$  can be expressed in terms of the Uhlmann fidelity \cite{uhlmann76,jozsa94},  which unveils the
 distinguishability between quantum states that are infinitesimally distant \cite{bures69}. The fidelity is defined as 
 \begin{equation}
  \mathcal{F}(\rho_1, \rho_2)=\left(\Tr\sqrt{\sqrt{\rho_1}\, \rho_{ 2} \, \sqrt{\rho_1}}\,\right)^2
  \label{eq:fidelity}
 \end{equation}
 and its connection to the QFI is expressed by the relation
 \cite{caves94,safranek17}
 \begin{equation}
 Q(\eta,t)=\lim_{\delta \eta \to 0} \frac{8\left(1-\sqrt{\mathcal{F}\big(\rho_{\eta}(t), \rho_{\eta+\delta\eta}(t)\big)}\,\right)}{\delta\eta^2}. \label{eq:qfifid}
 \end{equation}
In what follows we will also exploit an alternative, but equivalent, definition of the QFI, which may be introduced as follows: 
Given the time-local generator $\mathcal{L}(t)$ of the master equation
\begin{equation}
\frac{d\rho(t)}{dt}  = \mathcal{L}(t)[\rho(t)],
\end{equation}
the corresponding linear dynamical map is given by
\begin{align}
&\Lambda(t) = T_\leftarrow e^{\int_0^t d\tau \mathcal{L}(\tau)} \label{eq:dysonfull} \\
&=\sum_{k=0}^\infty \int_0^t dt_1 \mathcal{L}(t_1) \int_0^{t_1} dt_2 \mathcal{L}(t_2) \ldots \int_0^{t_k} dt_k\mathcal{L}(t_k) \nonumber. 
\end{align}
Given the orthonormal basis of operators $\{\tau_k\}_{k=0}^3 = \{\mathbb{1}/\sqrt{2},\sigma_x/\sqrt 2,\sigma_y/\sqrt 2,\sigma_z/\sqrt 2\}$, and  the Hilbert-Schmidt scalar product $\langle \xi,\chi \rangle \equiv
\Tr\llrr{\xi^\dagger \chi}$, any linear map $\mathcal{M}$ acting on a qubit state $\rho$ can be represented through a $4 \times 4$ matrix
\begin{equation}
\mathcal{M}[\rho] = \sum_{\alpha\beta=0}^3 D^\mathcal{M}_{\alpha\beta} \langle \tau_\beta, \rho \rangle \tau_\alpha \,\,\,\, \quad D_{\alpha \beta}^\mathcal{M} = \langle \tau_\alpha, \mathcal{M}[\tau_\beta] \rangle. \label{eq:tomatrix}
\end{equation}
Anlogously, a state $\rho$ can be written as a $4 \times 1$ column vector $\boldsymbol{\widetilde{r}} = (\langle \mathbb{1},\rho \rangle=1,\langle \sigma_x,\rho \rangle,\langle \sigma_y,\rho \rangle,\langle \sigma_z,\rho \rangle)^T$ containing the coefficients $\langle \tau_\alpha,\rho \rangle$ of the decomposition
\begin{equation}
\rho = \frac{1}{\sqrt{2}} \boldsymbol{\tau} \cdot\boldsymbol{\widetilde{r}} =\sum _{\alpha=0}^3\langle \tau_\alpha,\rho \rangle \tau_\alpha = \frac 1 2 \llrr{\mathbb{1}+\sum_{\alpha=x,y,z} \langle \sigma_\alpha, \rho \rangle},
\end{equation}
where the terms $\langle \sigma_\alpha,\rho \rangle,\ \alpha=x,y,z$ are the components of the Bloch vector associated to $\rho$. Given a completely positive trace preserving (CPTP) dynamical map $\Lambda(t)$, the most general form of the matrix $D^\Lambda$ associated with it is 
\begin{equation}
D^\Lambda = \left (
\begin{array}{c|c}
1 & \boldsymbol{0}^T \\
\hline
\boldsymbol{\nu} & V
\end{array}
\right ),
\label{Dlambda}
\end{equation}
where $\boldsymbol{0}^T$ is a 3-dimensional row vector, $\boldsymbol{\nu}$ is a real 3 dimensional column vector and $V$ is a $3 \times 3$ real matrix. By construction, therefore, the column $\boldsymbol{\nu}$ induces a state independent translation of the Bloch vector $\boldsymbol{r} = (\langle \sigma_x,\rho \rangle,\langle \sigma_y,\rho \rangle,\langle \sigma_z,\rho \rangle)^T$, whereas $V$ describes rotations, reflections and contraction of $\boldsymbol{r}$, so that
\begin{equation}
D^\Lambda \boldsymbol{\widetilde{r}} = {(1,\, \boldsymbol{\nu}+V \boldsymbol{r})^T}.
 \label{eq:affine}
\end{equation} 
As shown in \cite{nori13}, given an initial state $\widetilde{\boldsymbol{r}}(0)$, the quantum Fisher information associated to an unknown parameter $\eta$ of the $\eta$-dependent dynamical map $ \Lambda_\eta(t)=\Lambda(t) $ can be expressed as:
\begin{equation}
Q(\eta,t) = |\dot{D}^{\Lambda(t)} \widetilde{\boldsymbol{r}}(0)|^2 + \frac{({D}^{\Lambda(t)}\widetilde{\boldsymbol{r}}(0) \cdot \dot{D}^{\Lambda(t)} \widetilde{\boldsymbol{r}}(0))^2}{2-|{D}^{\Lambda(t)}\widetilde{\boldsymbol{r}}(0)|^2}, \label{eq:qfiNori}
\end{equation}
where $\dot{D}^{\Lambda(t)}$ indicates the derivative with respect to the parameter $\eta$.
\section{Weak-coupling limit}
The determination of the QFI requires the knowledge of the reduced system state $\rho_S(t) = \Tr_E[\rho_{SE}(t)]$, or equivalently of the dynamical map $\mathcal{M}(t)$ such that $\rho_S(t) = \mathcal{M}(t)\rho_S(0)$.
Such reduced state and dynamical map, as we mentioned before, are exactly analytically available in the spin-boson setting (Eqs. \ref{eq:completeHam}--\ref{eq:ohmic}) only for the specific case $\theta=\pi/2$, corresponding to a {pure dephasing} dynamics ($[H_S,H_I(\pi/2)]=0$)  \cite{breuer02}. For arbitrary values of $\theta$, instead, an analytically exact description of the evolved state $\rho_S(t)$ of the open system is, in general, not available.  In this section we study the short-time evolution of the probe qubit, and the corresponding behavior of the accuracy limits, as determined by the QFI, of the estimation of unknown environmental parameters.
A closed form of the master equation governing the dynamics of the probe qubit system interacting with a bosonic environment as described by \eref{eq:completeHam}, can be perturbatively derived in the weak coupling limit. By following the procedure described in \cite{Haase18}, involving a second-order time convolutionless (TCL) expansion, we end up with the master equation
\begin{align}
\frac{d\rho(t)}{dt} & = \mathcal{L}(t)[\rho(t)]=  -i\llrrq{H_S+H^{LS}(t),\rho(t)} \label{eq:tcl} \\
&+ \sum_{j,k = \pm,z} b_{kj}(t)\llrr{\sigma_k \rho(t) \sigma_j -\frac{1}{2} \left \{\sigma_j^\dagger \sigma_k,\rho(t) \right \}}, \nonumber \label{eq:tcl}
\end{align}
where $\sigma_{\pm} \stackrel{}{=} (\sigma_x \pm i \sigma_y)/2$.  Introducing the function
\begin{equation}
\Gamma(\xi,t) = \int_0^t d\tau e^{i \xi \tau} C(\tau),
\end{equation}
the time-dendent coefficients of the Lamb-shift Hamiltonian correction $H^{LS}(t)$ and of the dissipative part $b_{kj}(t)$ read (see eq. 36 of \cite{Haase18})
\begin{align}
b_{zz}(t) &= \frac{\sin^2(\theta)}{2} \Re[\Gamma(0,t)] \nonumber \\
b_{++}(t) &= \frac{\cos^2(\theta)}{2} \Re[\Gamma(-\omega_S,t)] \nonumber \\
b_{--}(t) &= \frac{\cos^2(\theta)}{2} \Re[\Gamma(\omega_S,t)] \nonumber \\
b_{+-}(t) &= b_{-+}^*(t)=\frac{\cos^2(\theta)}{4} (\Gamma(-\omega_S,t)+\Gamma^*(\omega_S,t)) \nonumber \\
b_{z+}(t) &= b_{+z}^*(t)=\frac{\sin(\theta)\cos(\theta)}{4} (\Gamma(0,t)+\Gamma^*(-\omega_S,t)) \nonumber\\
b_{z-}(t) &= b_{-z}^*(t)=\frac{\sin(\theta)\cos(\theta)}{4} (\Gamma(0,t)+\Gamma^*(\omega_S,t)) \nonumber\\
H_{11}(t) &= \frac{\cos^2(\theta)}{4} \Im[\Gamma(\omega_S,t)] \nonumber \\
H_{10}(t) &= H_{01}^*(t) = \frac{-i \sin(\theta)\cos(\theta)}{4}  \nonumber \\
& \left(\Re[\Gamma(0,t)] - \frac{1}{2}  (\Gamma^*(-\omega_S,t)+\Gamma(\omega_S(t))\right) \nonumber\\
H_{11}(t) &= \frac{\cos^2(\theta)}{4} \Im[\Gamma(-\omega_S,t)], \label{eq:tclcoeff}
\end{align}
where $\Re[\cdot], \Im[\cdot]$ indicate the real resp. imaginary part, $c^*$ the complex conjugate of $c$ and $H_{ij}^{LS}(t) = \bra{i} H^{LS}(t)\ket{j}, i,j=0,1$.  Our aim here is to obtain the short-time solution of the master equation \eref{eq:tcl}. To this end, it is sufficient to consider only the terms of $D^{{\Lambda}(t)}$ up to some order $k$ in $t$.  

In order to get an insight into the dependence of the QFI, in the very initial phase of the dynamics, on the initial condition $\rho_S(0)$ and on the interaction angle $\theta$, we start by considering the ($2$-nd order) Dyson expansion of $D^{\Lambda(t)}$ with only terms up to $t^2$. The resulting super-operator matrix $D^{\Lambda(t)}_{(2)}$ reads
\begin{widetext}
\begin{equation}
D^{\Lambda(t)}_{(2)}=\left(
\begin{array}{cccc}
 1 & 0 & 0 & 0 \\
 0 & 1- \frac{t^2}{2}(\zeta(0) \sin^2(\theta) +\omega_S^2) & -t
   \omega_S & \frac{1}{4}
   \zeta (0) t^2 \sin (2 \theta ) \\
 0 & t \omega_S &
   1-\frac{t^2}{2}  \left(\zeta
   (0)+\omega_S^2\right) &
   0 \\
 0 & \frac{t^2}{4} \zeta (0) \sin
   (2 \theta ) & 0 & 1-\frac{t^2}{2}
   \zeta (0)  \cos ^2(\theta ) \\
\end{array}
\right),
\end{equation}
\end{widetext}
where $\zeta(n)$ indicates the $n$-th moment of the spectral density, i.e.
\begin{align}
\zeta(n) & = \frac{1}{i^n} \left .\frac{d^n}{dt^n}C(t) \right |_{t\to 0}.
\end{align}
Since the QFI is convex, we restrict our attention to pure initial states; moreover, for the sake of simplicity, we restrict the initial states to lie in the $x-z$ plane, so that the initial condition can be parametrized by a single angle $\alpha$ as 
\begin{equation}
\bs{r}_0(\alpha) = (\cos(\alpha),0,\sin(\alpha))^T.
\label{initialState}
\end{equation}
 By exploiting \eref{eq:qfiNori}, it is easy to determine the QFI for the estimation of an arbitrary environment parameter $\eta$. The leading order term is proportional to $t^2$ and reads
\begin{equation}
{Q}_{(2)}(\eta,t) = \frac {t^2} {4} \sin^2(\alpha-\theta) \frac{(\partial_\eta \zeta(0))^2}{\zeta(0)},
\label{eq:qfi2Nori}
\end{equation}
where we indicate by ${Q}_{(k)}$  the QFI corresponding to $D^{\Lambda(t)}_{(k)}$,  i.e. the matrix obtained by keeping the terms up to $t^k$ of the Dyson expansion \eref{eq:dysonfull}. 
For arbitrarily chosen, but fixed, environmental parameters the steepest increase of ${Q}(\eta,t)$, at short times, is thus provided by the choices $\alpha-\theta =  \pi/2+k \pi,\ k \in \mathbb{Z}$. If a pure dephasing dynamics ($\theta=\pi/2$) is considered, for example, the initial states maximising the initial increase in the QFI \eref{eq:qfi2Nori} correspond to $\alpha =  k \pi,\ k \in \mathbb{Z}$, namely the eigenstates of $\sigma_x$, which are already known to be the optimal ones in this case. In the presence of a purely transverse system-environment interaction ($\theta = 0$), instead, the initial state maximising the initial growth of the QFI  is given by the choice  $\alpha= \pi/2+k \pi, \ k \in \mathbb{Z}$, i.e. the eigenstates of $\sigma_z$. We moreover point out that different combinations of $\alpha$ and $\theta$ resulting in the same value $\alpha-\theta$ will lead to the same initial increase of the QFI.

We notice that \eref{eq:qfi2Nori} is independent on the system frequency $\omega_S$; such dependence emerges only if higher order Dyson expansions $D_{(k)}^{\Lambda(t)}$ and the corresponding $\text{Q}_{(k)}(\eta,t)$ are considered. This means that $\omega_S$ dependent terms can contribute to the QFI, and thus be used as another control parameter of the probe qubit, only for sufficiently large times, or stronger system-enviroment couplings.
This can be seen by analyzing  the matrix form
 for the generator $\mathcal{L}(t)$, derived by using 
\begin{equation}
D^{\mathcal{L}(t)}_{(3)} = \left (
\begin{array}{c|c}
0 & \boldsymbol{0} \\
\hline
\boldsymbol{\mu}_{(3)}(t) & W_{(3)}(t)
\end{array}
 \right),
\end{equation}
with $\boldsymbol{0}$ the three-dimensional zero vector, 
\begin{equation}
\boldsymbol{\mu}_{(3)}(t) = \frac{t^3 \zeta(1) \omega_S}{6} \left (\sin (2 \theta),0, -2 \cos^2(\theta) \right )^T, \label{eq:firstcol3}
\end{equation}
  and the matrix $W_{(3)}(t)$ is fully defined in Appendix \ref{appa}.
 {$D^{\mathcal{L}(t)}_{(3)} $ reveals that such dependence on $\omega_S$ appears indeed only for $k \geq 3$ (or $k\geq 4$ if $D^{\Lambda(t)}_{(k)}$ is  considered).
 
 Moreover, {it is interesting to notice  from Eq. \eqref{Dlambda} that} the dynamical map $\Lambda(t)$ loses its unital character, namely $\Lambda(t)[ \mathbb{1}] \neq \mathbb{1}$, but for  $\theta = \pi/2 + k \pi,\ k=1,2,\ldots$, i.e. for pure dephasing dynamics. 
  Since the translation term $\bs{\mu}(t)$ in the generator $D^{\mathcal{L}(t)}_{(3)}$ is proportional to $t^3$, however, the lowest order contribution to the translational part $\boldsymbol{\nu}(t)$ (see \eref{eq:affine}) to the dynamics is of order $t^4$. For very short times, therefore, the translations of the Bloch vector will be negligible, and the map will be approximately unital. On the other side, this fact suggests that the dynamics of the Bloch vector over longer times, or in the presence of a stronger coupling to the environment, will be affected by environment-dependent translations; this can affect the dependence of the probe state on the environmental parameters, and lead to an increase of the QFI related to the estimation of these latter. We moreover observe that, by avoiding the high temperature limit $\beta \to 0$ used in \cite{Haase18} our setting allows to address the estimation of system or environmental parameters in any temperature range.

{Beside providing an analytic insight on some of the features of the dynamical map, the Dyson expansions $D^{\Lambda(t)}_{(k)}$ turn out to be  most useful for the numerical analysis of the dependence of the QFI on the interaction and initial state parameters in the weak-coupling/short-time regime we are discussing here. For the computation of the QFI $Q(\eta,t)$ by means of \eref{eq:qfifid} the evolved states $\rho_\eta(t)$ and $\rho_{\eta+\delta\eta}(t)$  are needed. Such states can be determined by numerical integration of the TCL master equation \eref{eq:tcl}. However, for the small increments $\delta\eta$ required for good finite-difference approximations of the infinitesimal increment limit $\delta\eta \to 0$, numerical instabilities can arise. In fact, $\rho_\eta(t)$ and $\rho_{\eta+\delta\eta}(t)$ start from the same initial state, and at very short times/weak coupling, the difference between the evolved states is typically very small. It is easy to check that such instabilities are much more pronounced, and appear over longer time-intervals, in the presence of energy-exchange type of interaction ($\theta = 0$ in our setting) alone: energy-exchange processes typically occur on longer times (see \fref{fig:fidelity}(b)).
}

{In what follows we will therefore adopt a different approach and determine the evolution of the probe qubit by means of $D_{(7)}^{\Lambda(t)}$. On the one hand, it provides excellent approximation of the dynamical map determined by the TCL master equation up to $t\approx 0.4$, as exemplified in  \Fref{fig:fidelity}(a); on the other, it allows for an analytic derivation of the QFI by means of \eref{eq:qfiNori}.}

\begin{figure}
\subfigure[]{\includegraphics[width=0.8\columnwidth]{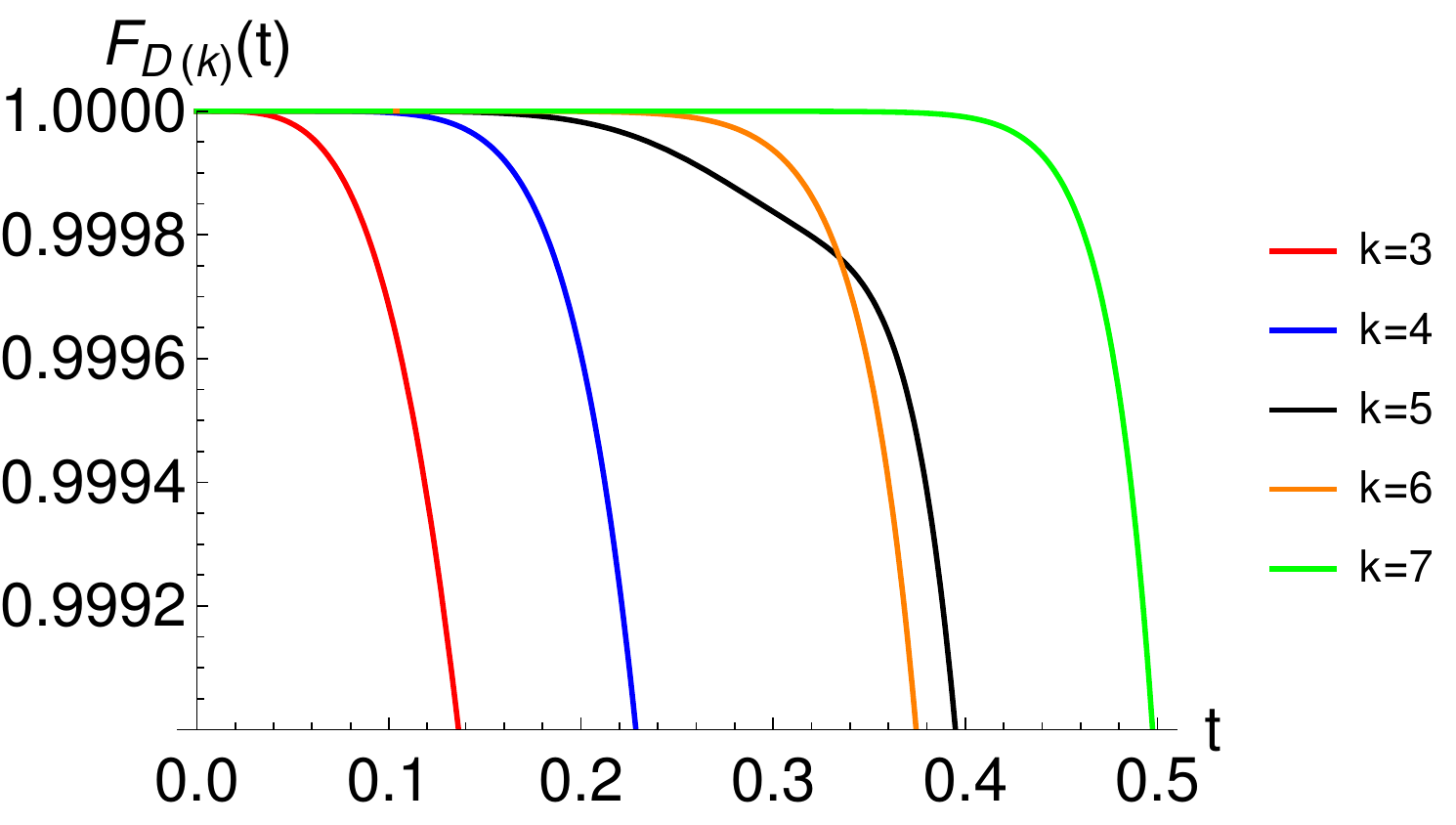}}
\subfigure[]{\includegraphics[width=0.71\columnwidth]{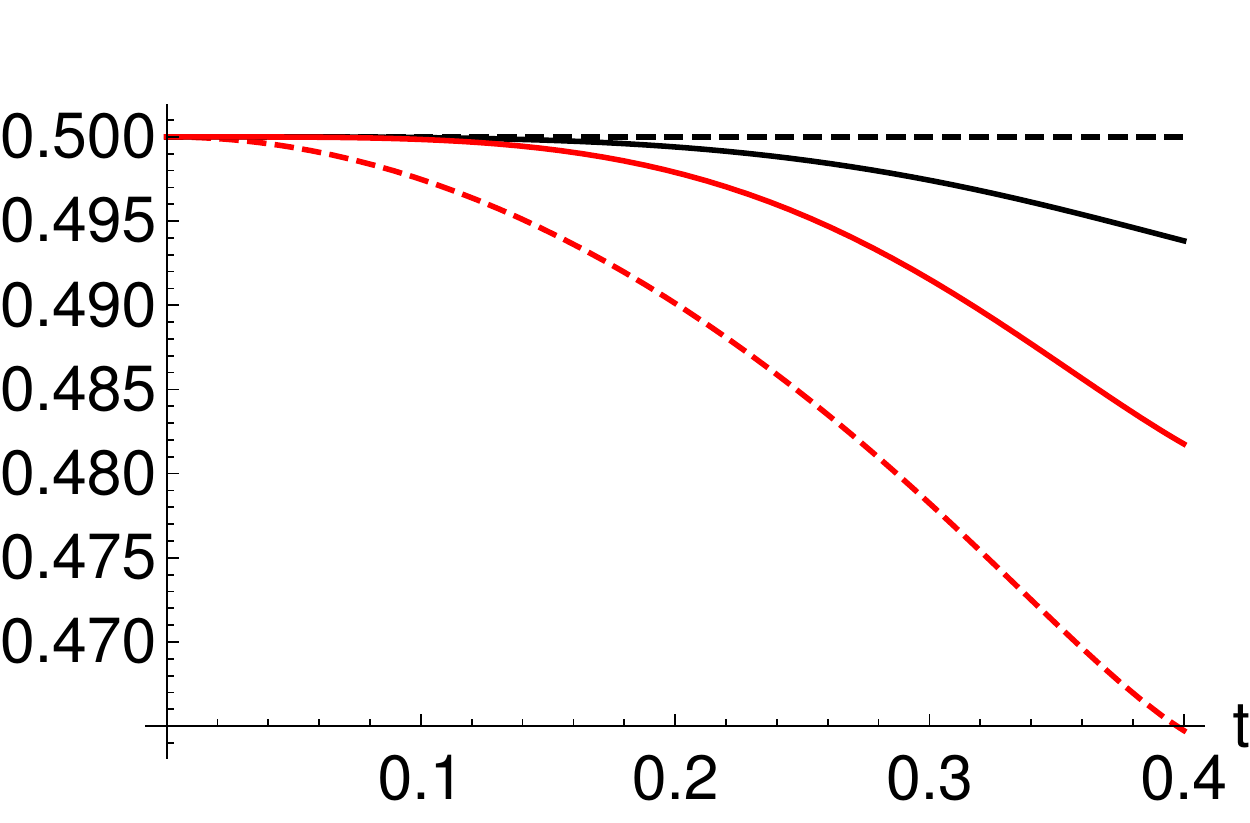}}
{\caption{ \label{fig:fidelity} {In both frames:  $\lambda=1,\ s=1, T=0.1,\omega_S=5$ (remind that $\omega_c=1$), initial state $\rho_S(0)=\ketbra{+}{+}$. Panel (a): The fidelity \eref{eq:fidelity} between the numerical solution $\rho_S^{\text{TCL}}(t)=\Lambda(t) \rho_S(0)$ of the TCL master equation \eref{eq:tcl} and $\rho_S^{D(k)}(t) = D_{(k)}^{\Lambda(t)} \rho_S(0)$  as a function of time for different values of the time-expansion order $k$. Panel (b): The evolution of the excited state population (black) and of the absolute value of the coherence (red) when the initial state $\rho_S(0)$ is evolved under a pure dephasing dynamics ( $\theta=\pi/2$, dashed) and a completely transverse dynamics ($\theta=0$, solid).}}}
\end{figure}
\begin{figure*}[t]
\center
\subfigure[]{\includegraphics[width=0.55\columnwidth]{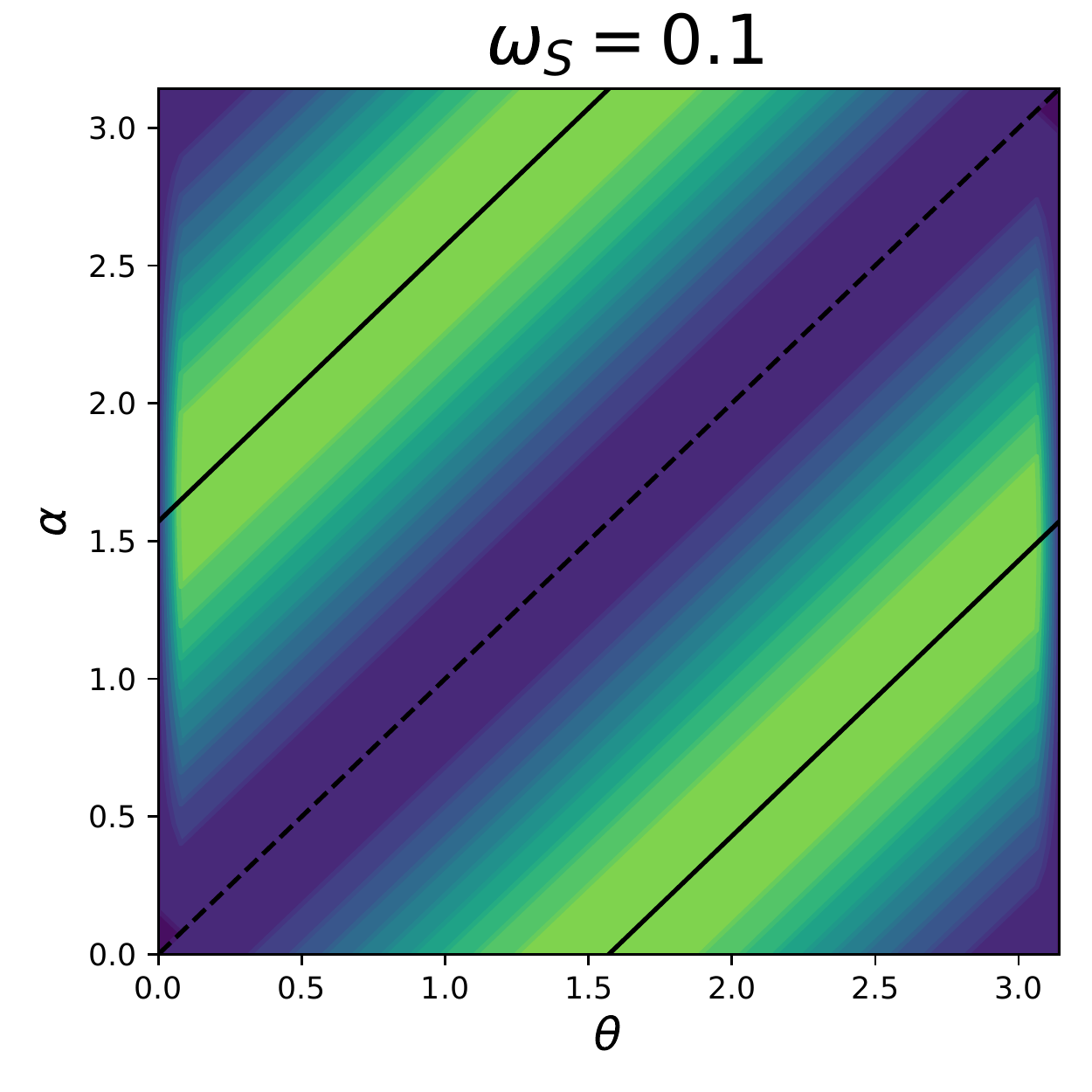}}
\subfigure[]{\includegraphics[width=0.55\columnwidth]{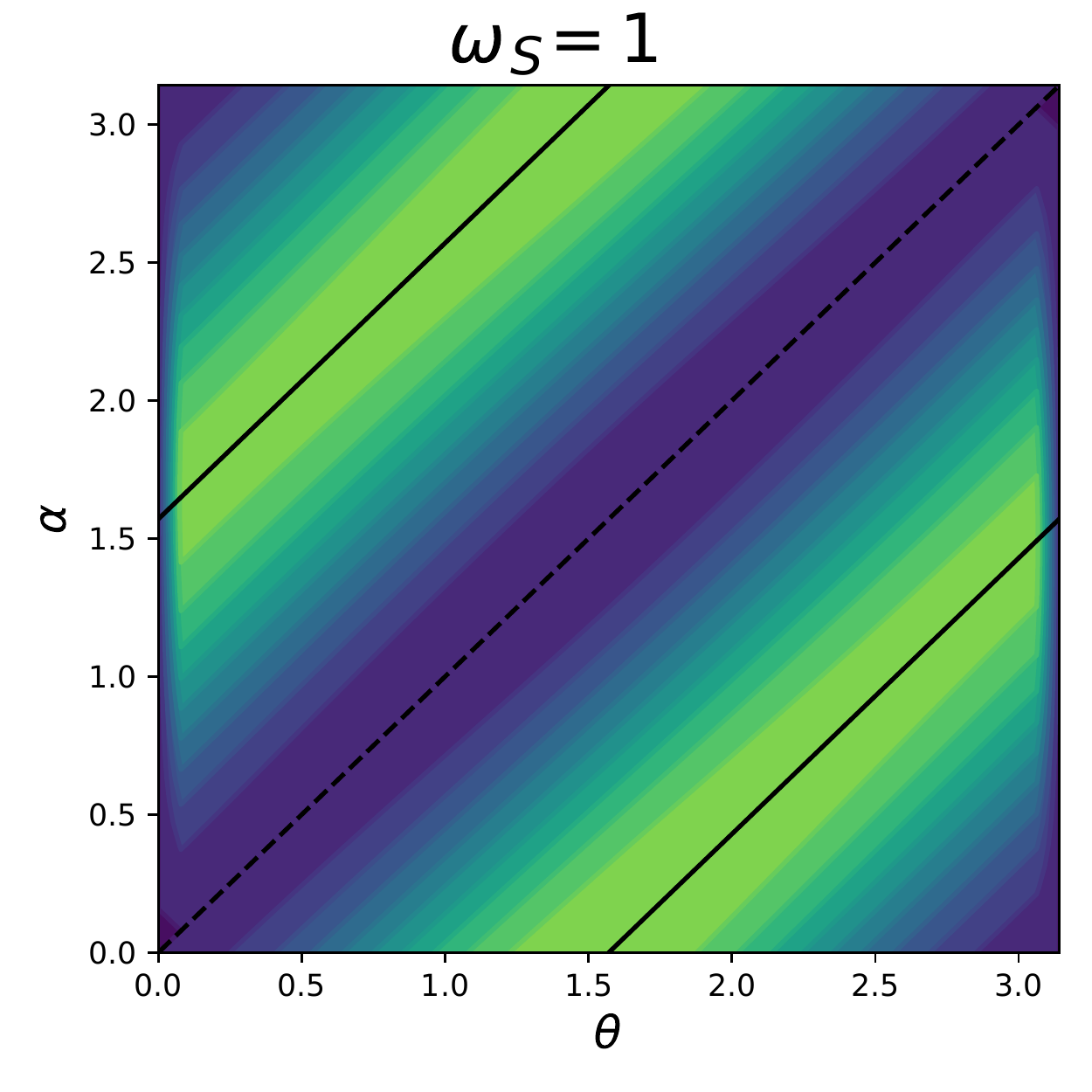}}
\subfigure[]{\includegraphics[width=0.66\columnwidth]{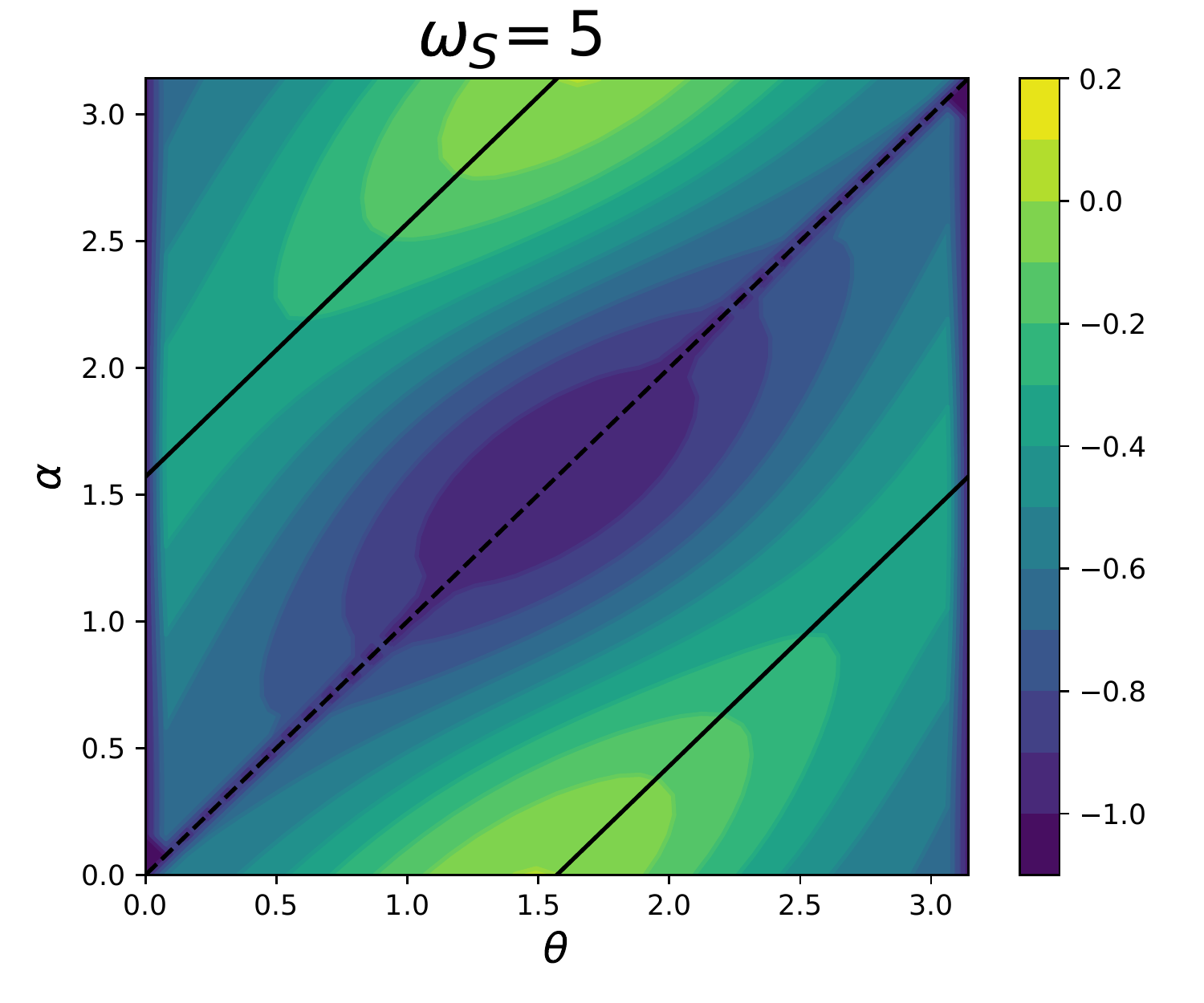}}\\
\subfigure[]{\includegraphics[width=0.55\columnwidth]{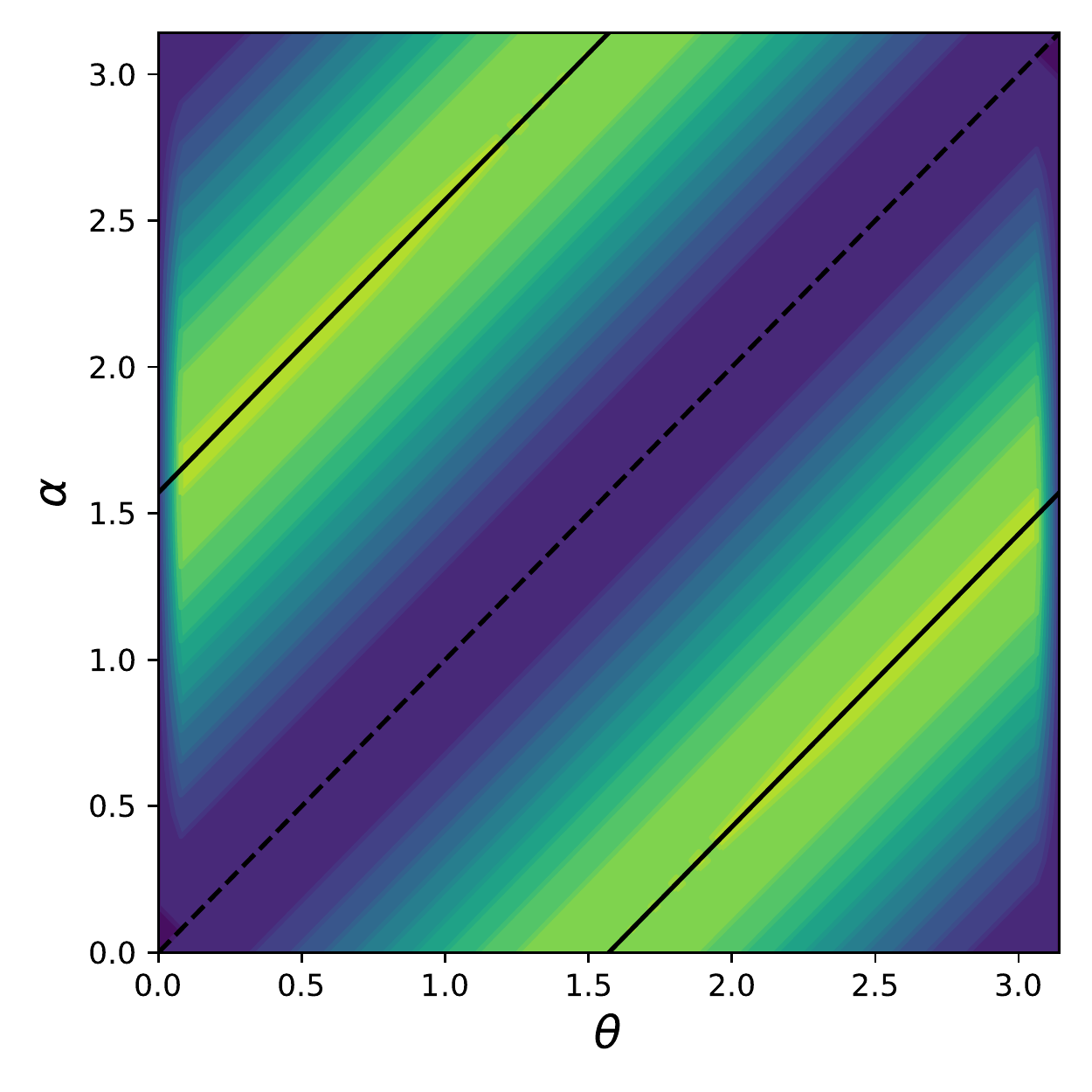}}
\subfigure[]{\includegraphics[width=0.55\columnwidth]{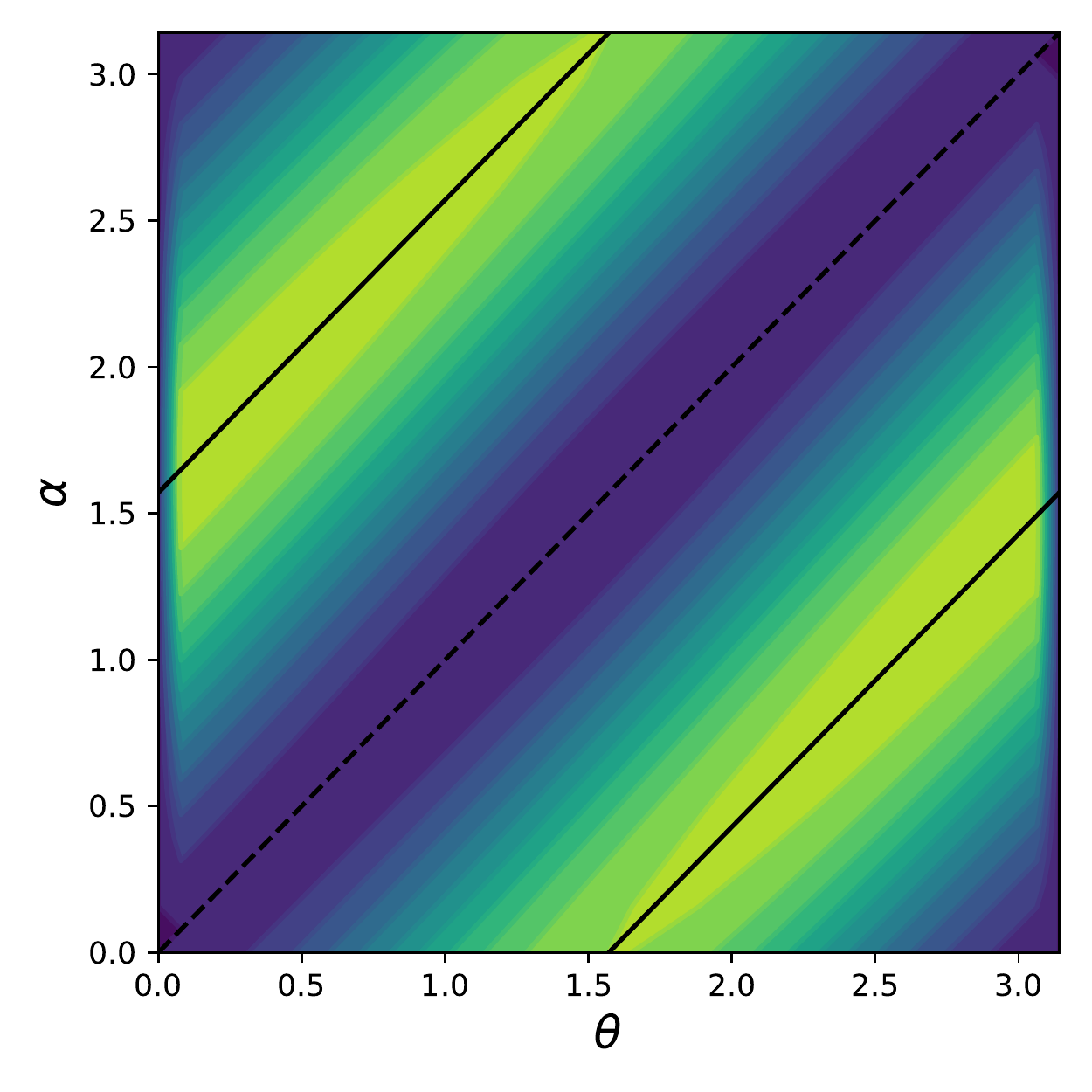}}
\subfigure[]{\includegraphics[width=0.66\columnwidth]{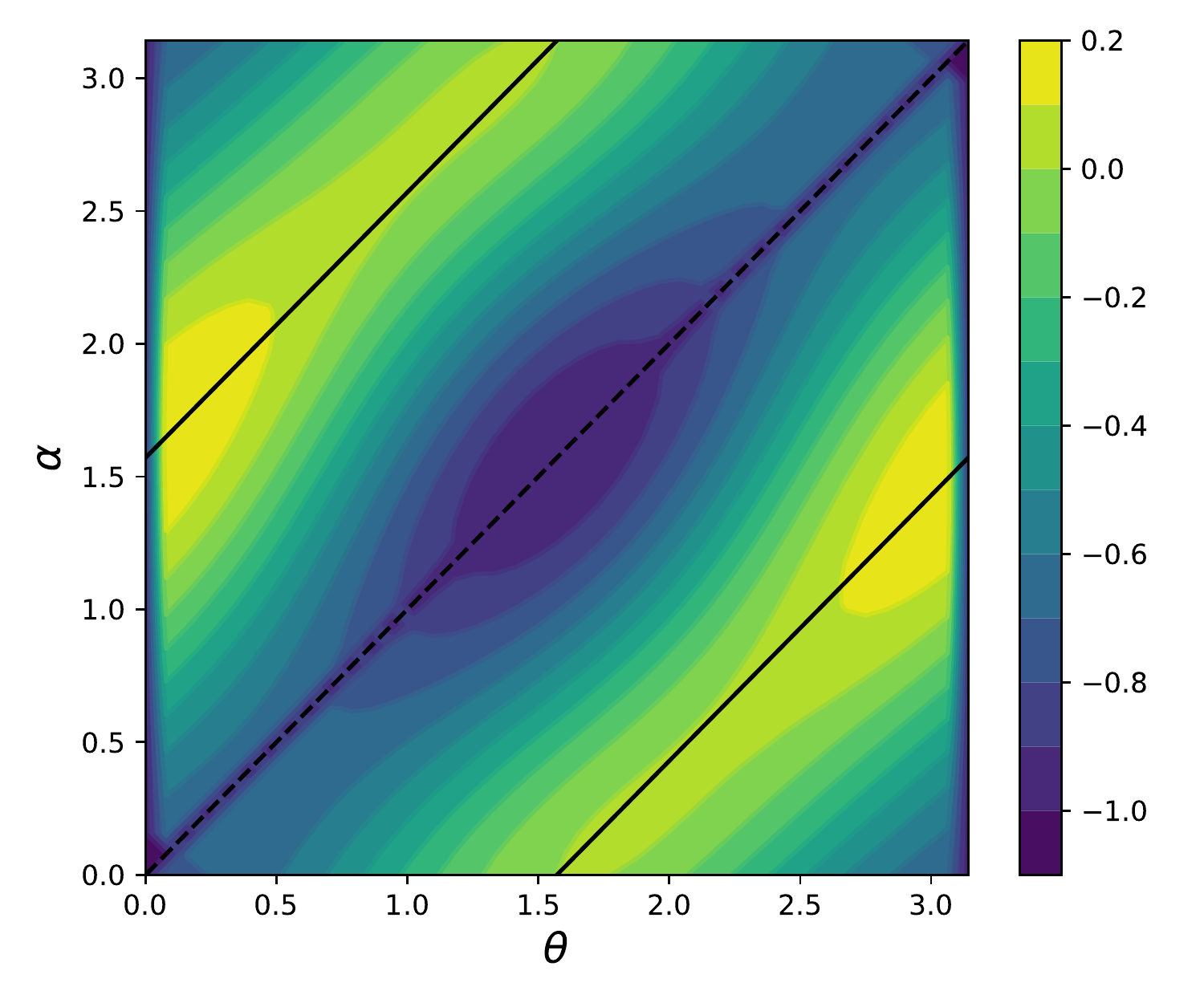}}
{\caption{\label{fig:mapFigs} All frames: $\lambda=1,\ s=1,\ T=0.07 \ (\beta = 1/T \approx 14.3),\ t=0.35$. Panels (a)-(c): 
Temperature estimation with $R_t(\beta,\theta,\alpha)$ for $\omega_S=0.1, 1$ and $5$ (remind that $\omega_c=1$). Panels (d)-(f): Cutoff frequency estimation with  $R_t(\omega_c,\theta,\alpha)$ for the same three values of $\omega_S$. Black solid lines, corresponding to the points $(\theta,\alpha = \theta \pm\pi/2)$, and black dashed lines, corresponding to the points $(\theta,\alpha=\theta)$,  are used as a guide to the eye to locate the parameters regions corresponding respectively to the larger and smaller values of the ratio $R_t(\eta,\theta,\alpha)$.}}
\label{shortTimesRatio}
\end{figure*}

In order to quantify the optimality of pure dephasing we introduce the ratio
\begin{align}
R_t(\eta,\theta,\alpha)=\frac{{Q}^{\theta,\alpha,\omega_S}(\eta,t) -{Q}^{\frac{\pi}{2},0,\omega_S}(\eta,t)}{{Q}^{\frac{\pi}{2},0,\omega_S}(\eta,t)},
\end{align}
namely the relative difference between the QFI determined by the evolution of the probe system having free dynamics determined by $H_S = \frac{1}{2}\omega_S \sigma_z$  initially in the state  $\bs{r}_0(\alpha)$ and interaction Hamiltonian $H_I(\theta)$ and the QFI at  the same time provided by a pure dephasing dynamics of the probe qubit starting from the (dephasing-optimal) initial state $\rho_S(0)=\ketbra{+}{+}$, as a figure of merit for the estimation of the environmental parameter $\eta$.
We have $R>0$ when a strategy outperform the performance of dephasing.

We apply  our setting to the study of the QFI associated with the {short-time} estimation of the bosonic bath (inverse) temperature $T$ ($\beta$) and of the cutoff frequency $\omega_c$, 
i.e. $Q(\beta,t)$ and $Q(\omega_c,t)$, respectively. An identical procedure  can be clearly applied to other environmental parameters, such as $\lambda$ and $s$. 
 
Our calculations  show that a pure dephasing dynamics acting on the initial state $\bs{r}_{0}(0)$ is optimal {for temperature estimation}: other combinations of the interaction angle $\theta$ and of the initial state angle $\alpha$ lead to smaller $Q(\beta,t)$ \footnote{The optimality of pure dephasing dynamics together with the choice $\alpha=0$ is not clearly visible in frame (a) of Fig. 2: the points $(\theta=\pi/2, \alpha=0)$  and $(\theta=\pi/2, \alpha=\pi)$ are indeed the only ones where $Q(\beta,t)=1$; the other points along the black lines always correspond, in this frame,  to slightly smaller values of $Q(\beta,t)$.}. 
{This can be seen in  \Fref{shortTimesRatio}(a)-(c), which }shows the ratio $R_t(\beta,\theta,\alpha)$ at  time $t=0.35$. The behavior is qualitatively the same at any $t \leq 0.35$ and
for different values of the Ohmicity parameter $s$ and whenever $\omega_c \gtrsim \omega_S $. 
As clearly visible in frames (a)-(c) of \Fref{shortTimesRatio} (see solid and dashed lines), initial states ``orthogonal'' to the interaction angle ($\alpha = \theta \pm \pi/2$) lead in general to higher values of the QFI. This is particularly evident in the case $\omega_S \ll \omega_c$ (\Fref{shortTimesRatio}(a)) where any choice $\alpha=\theta+\pi/2$ leads to the same $Q(\beta,t)$, as already predicted by the short-time expansion \eqref{eq:qfi2Nori}. As the system frequency $\omega_s$ is increased, instead, only the  dephasing  with initial state angle $\alpha=0$ is optimal.The QFI is instead minimized when the initial state is parallel to the interaction angle ($\alpha = \theta$).

The situation is different when $Q(\omega_c,t)$ is considered (frames (d)-(f) of  \Fref{shortTimesRatio}). For $\omega_S \geq \omega_c$ a purely transverse interaction term and an initial condition parallel to the $z$ axis outperforms pure dephasing at the considered time $t=0.35$ (see \Fref{shortTimesRatio}(f)) and, as we will see in the next section, for longer times. For shorter times, instead, pure dephasing dynamics with the initial state corresponding to $\alpha=0$ remains optimal (not shown). This suggests that energy exchanges between the system and the environment, which typically occur on longer times, see \Fref{fig:fidelity}(b), can provide additional information on the bath cutoff energy $\omega_c$. As in the case of temperature estimation, initial conditions orthogonal to the interaction direction lead to larger values of the QFI, whereas initial conditions parallel to the interaction direction correspond to smaller values of the QFI.

\section{Long times/arbitrary coupling}
The analysis of the previous section was
limited to the weak-coupling regime and short times.
Intuition suggests, on the other hand, that a stronger or longer interaction
 of the probe qubit with the environment could allow for a larger information gain on the environmental features, and therefore to an increase of the ultimate precision of the estimation of environmental parameters. 
 Moreover,  by extending the interaction time, the exchange of energy between system and environment which typically occur on time-scales much longer than the one characteristic of pure dephasing, 
 can become more relevant. An indication in this direction was already provided by the behavior of $R(\omega_c,\theta,\alpha)$ for $\omega_S \gg \omega_c$ (see \Fref{shortTimesRatio}(f)).

As well known, an analytic solution of the spin-boson model \eref{eq:completeHam} for arbitrary times and coupling strength is however available only for pure dephasing. For general directions of the system-environment interaction term a numerical solution is needed.  

In this section, we explore, by  numerical means, the behavior of the QFI associated to the temperature ($\eta=T$) and cutoff  frequency ($\eta=\omega_c$) estimation for different directions of the interaction term $H_I(\theta)$ and initial states $r_0(\alpha)$. More specifically, we use the T-TEDOPA \cite{tamascelli19} method in order to determine $\rho_S(t)$ in a numerically exact way. 
We refer the reader to  Appendix \ref{tedopaApp}  for a streamlined description of T-TEDOPA and all the details needed to reproduce our results and to extend the analysis to other environmental parameters not discussed here, such as the overall coupling $\lambda$ and the Ohmicity $s$.

In our numerical analysis we limited ourselves to consider the three initial states corresponding to $\alpha = 0$, $\alpha=\pi/4$ and $\alpha=\pi/2$ and  interaction angles $\theta \in [0,\pi/2]$.
The considered initial states and interaction angles are enough to see a rich variety of behaviours of the QFI, and in particular to show that a pure dephasig interaction is never  optimal for the estimation of bath parameters 
if we consider dynamics over long times, allowing for system-environment energy exchange processes to occur.

Clearly enough, the numerical approach does not allow for an analytic derivation of the QFI, as equivalently defined in equations \eref{eq:qfifid} or \eref{eq:qfiNori}, which would require a truly infinitesimal $\delta \eta$. We instead adopted a finite-difference approach: we derived, for any considered initial condition and interaction angle, the matrices $\rho_\eta(t)$ and $\rho_{\eta+\delta \eta}(t)$ by changing the estimated parameter in the spectral density by $\delta \eta$.  In what follows we set $\delta \eta = 10^{-4}$, which provides converged values of the QFI (smaller values of $\delta\eta$ lead to the same result). It is worth noting here that we are interested in the behaviour of the QFI over times much longer  than those considered in the previous section, so that such finite-different approach can be safely adopted: while the numerical instabilities due to the closeness, at short times, of the states $\rho_\eta(t)$ and $\rho_{\eta+\delta \eta}(t)$  are still there, they do not affect the comptuation of the QFI at longer times, where, in general, the distance between the two evolved states is larger.

\begin{figure*}[t]
\begin{minipage}{1. \textwidth}
\includegraphics[width=.95\columnwidth]{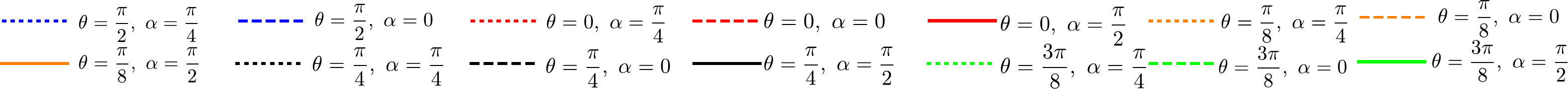}
\end{minipage}
\begin{minipage}{1.\textwidth}
\subfigure[]{\includegraphics[width=0.32\columnwidth]{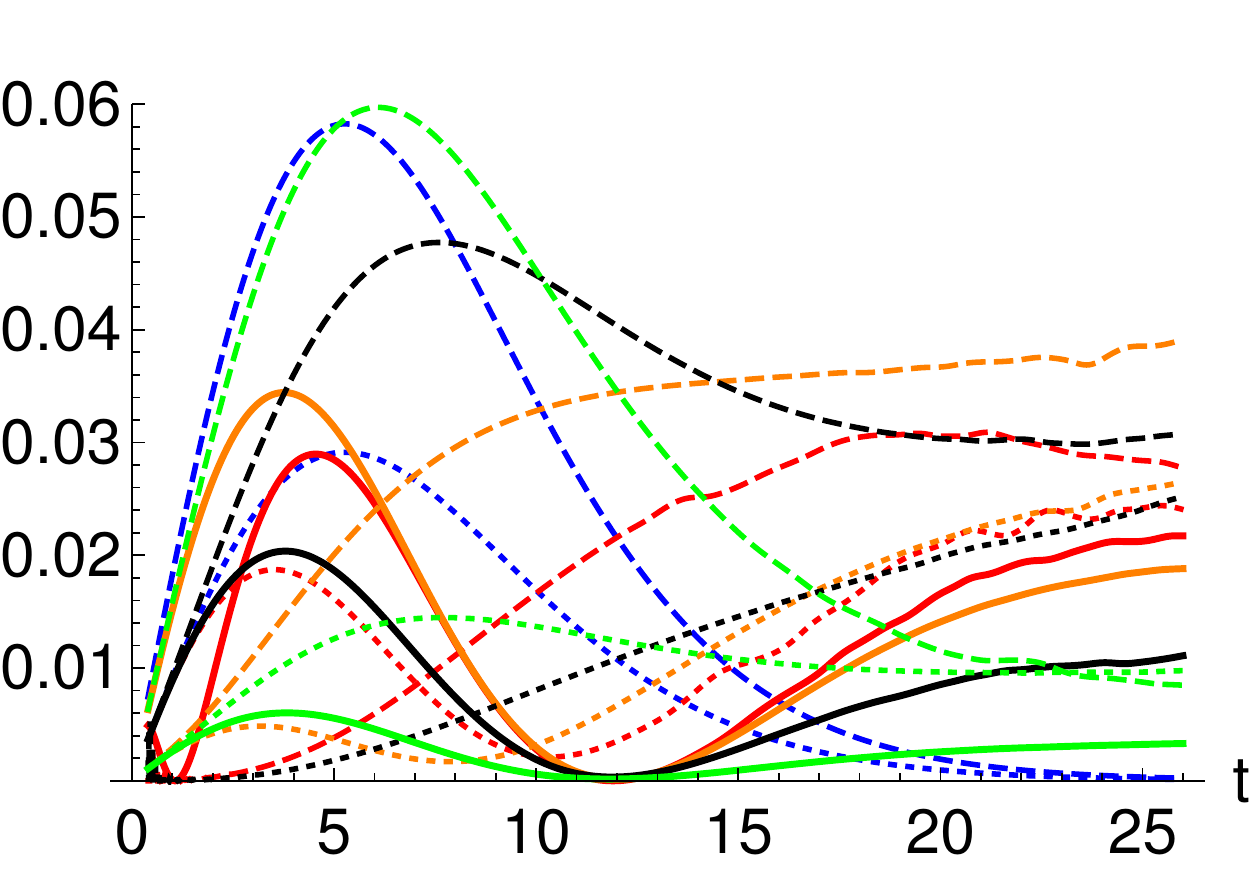}}
\subfigure[]{\includegraphics[width=0.33\columnwidth]{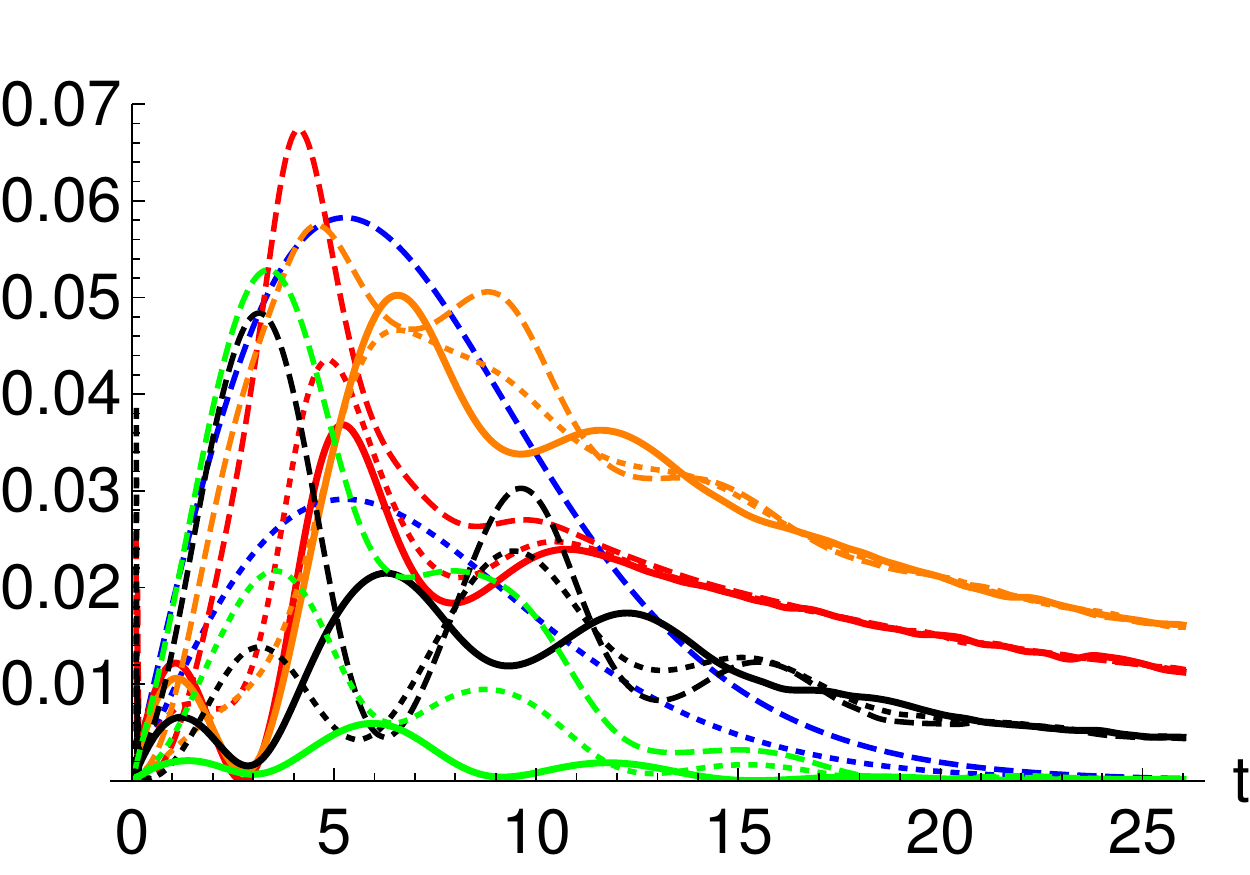}}
\subfigure[]{\includegraphics[width=0.31\columnwidth]{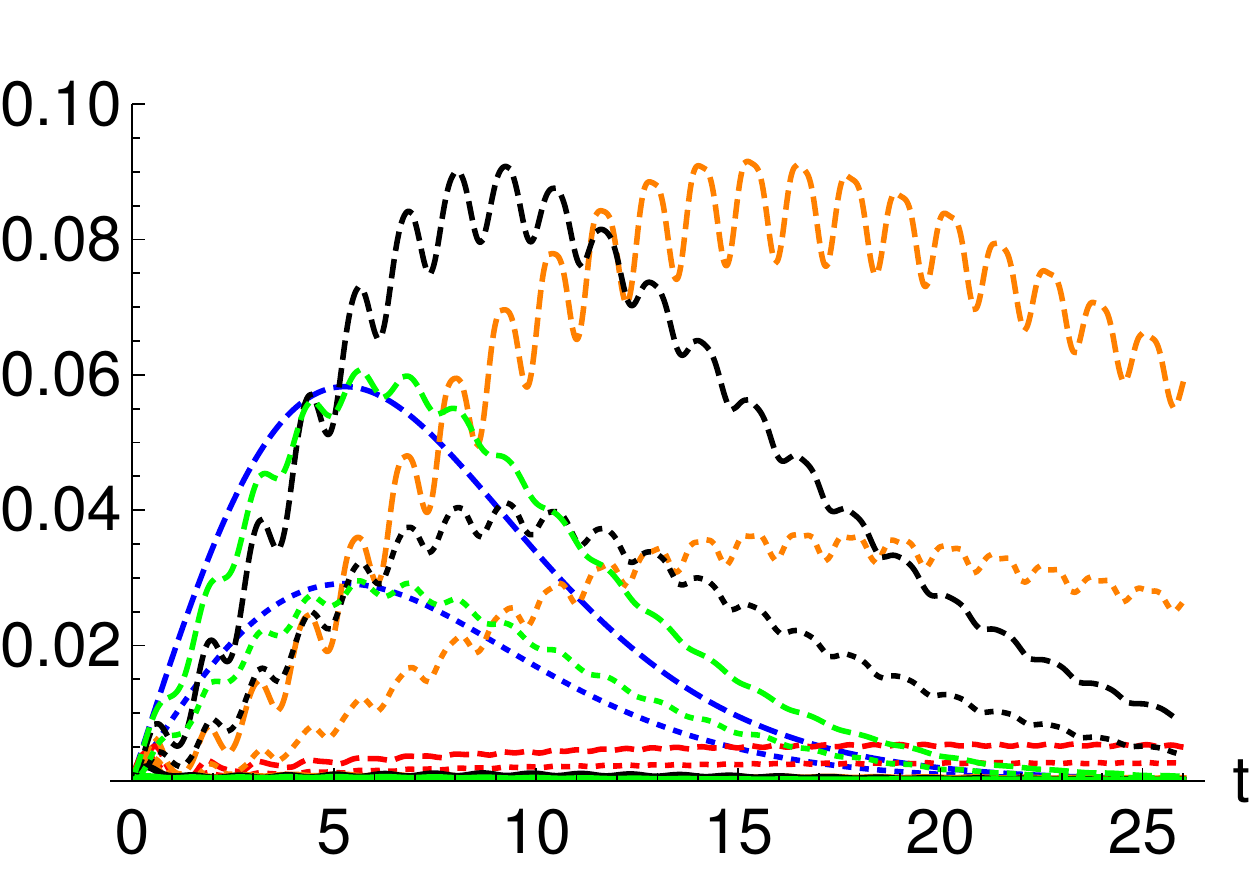}}
\\
\subfigure[]{\includegraphics[width=0.32\columnwidth]{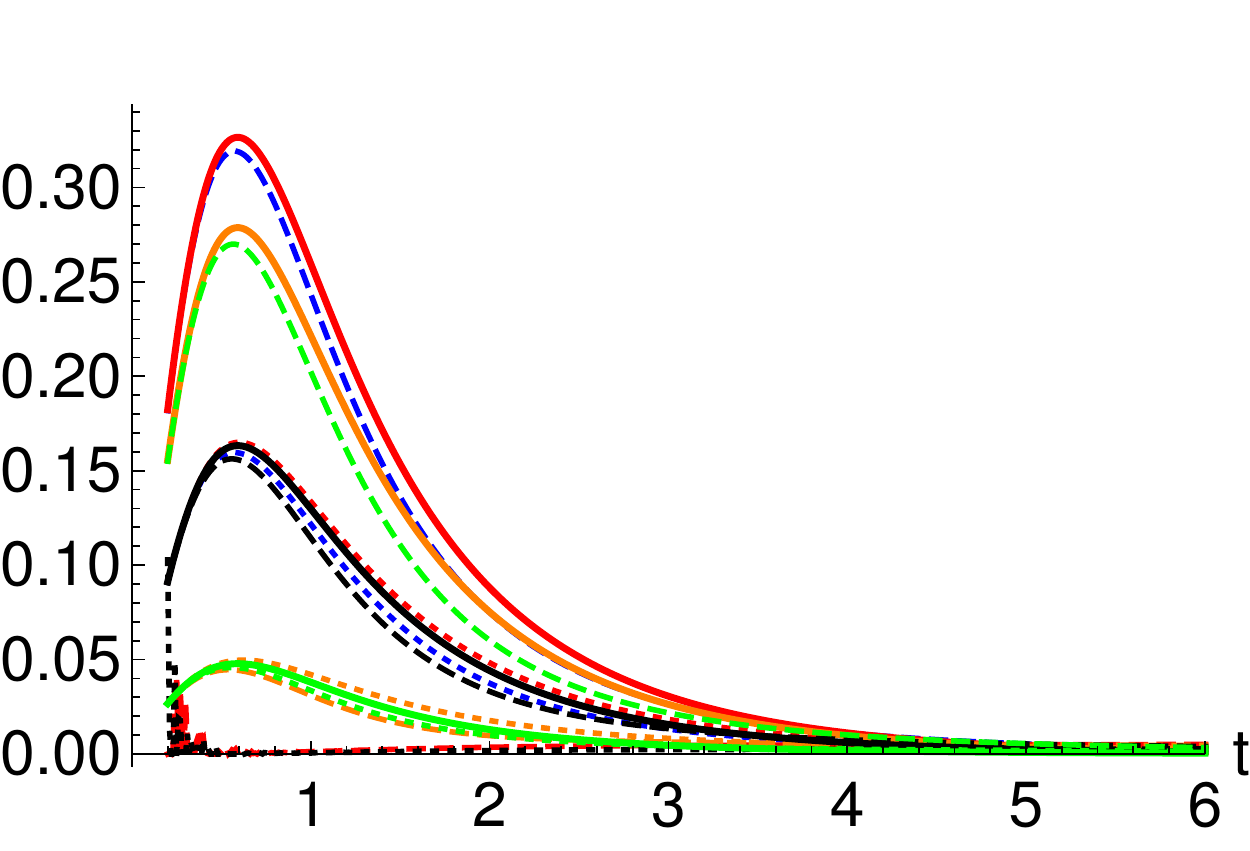}}
\subfigure[]{\includegraphics[width=0.33\columnwidth]{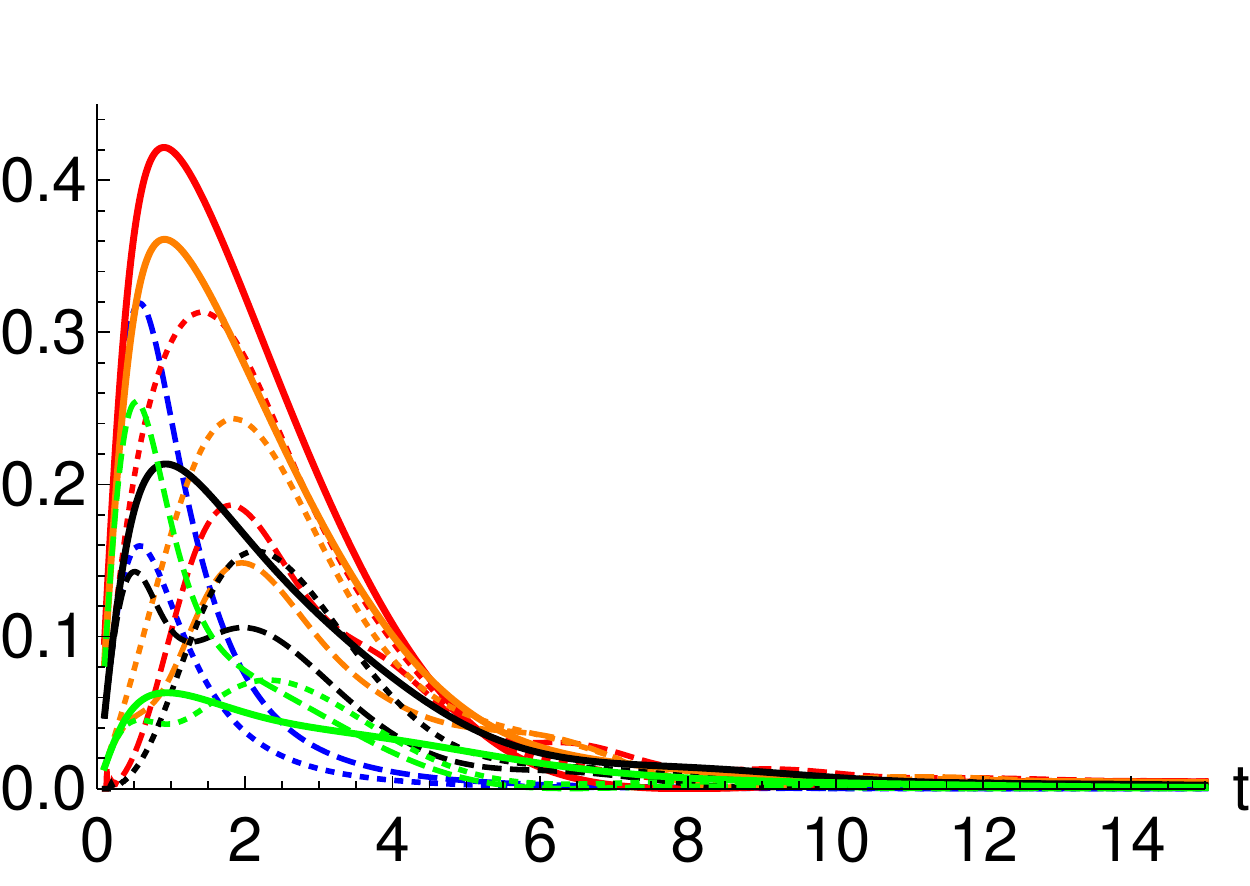}}
\subfigure[]{\includegraphics[width=0.31\columnwidth]{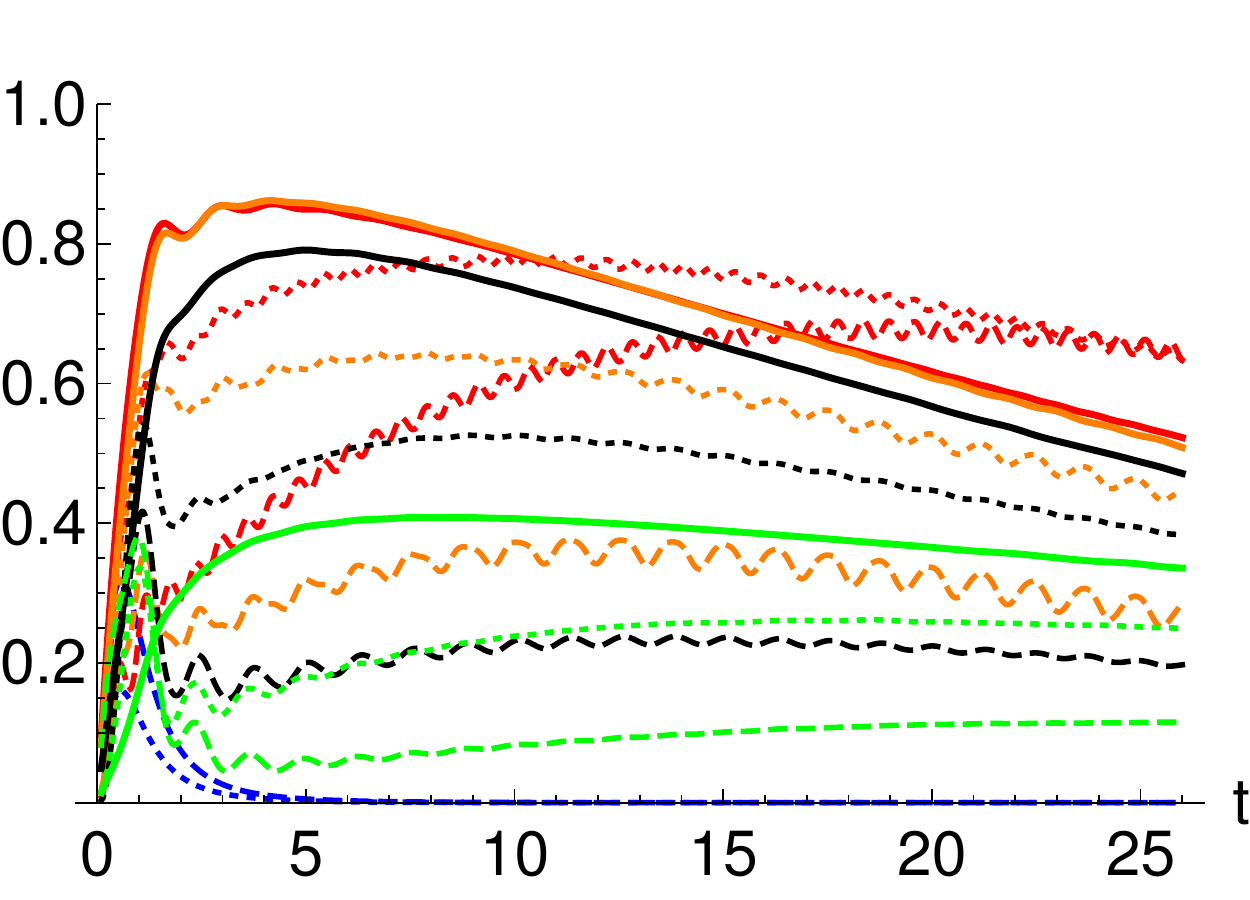}}
\end{minipage}
{\caption{\label{fig:tedopaFig} All plots: $q(\eta,t) = {Q}(\eta,t)/t$ as a function of $t$,  $\lambda=1,\ s=1,\ T=0.07$ (remind that $\omega_c=1$). Frames (a)-(c) Temperature estimation ($\eta=\beta$) for (a) $\omega_S=0.1$, (b) $\omega_S=1$,(c) $\omega_S=5$. Frames (d)-(f): Cutoff frequency estimation ($\eta = \omega_C$) for (d) $\omega_S=0.1$, (e) $\omega_S=1$, (f) $\omega_S=5$.  }}
\end{figure*}
{Instead of looking directly at the quantum Fisher information,  
we analyze the behavior of the QFI rescaled with time, i.e.
\begin{equation}
q(\eta,t)=\frac{Q(\eta,t)}{t}.
\end{equation}
In a metrological context, where time is a resource, it is important to be able to perform the
measurements in a short time or, otherwise stated, it is important to 
have a large repetition rate for the measurement.
The quantity $q(\eta,t)$ takes into account the fact that a large QFI at long times 
may be less advantageous with respect to a lower QFI at shorter times. 
High  values for the quantity $q(\eta,t)$ thus indicates a large  information gain
in a metrological sense.
}
Figure \ref{fig:tedopaFig}(a)-(c) show the behavior of  $q(\beta,t)$ {in time, 
for different combination of the $(\theta,\alpha)$ angles and}
 for different values of the system frequency $\omega_S$. 
The time-evolution of the {rate $q$} for pure dephasing dynamics for the optimal initial state corresponding to the choice $\alpha=0$ is clearly independent of the system frequency $\omega_S$; it exhibits a maximum at $t \approx 5$ and steadily decreases, getting close to zero around  $t=25$. For other values of $\theta$ the behavior of $q(\beta,t)$ shows a strong dependence on $\omega_S$ and there are combinations of interaction angle $\theta$, initial state angle $\alpha$ and times leading to
higher values of  $q(\beta,t)$  than the one achievable with pure dephasing dynamics. This is particularly evident if, for example, $\omega_S = 5$ is considered. It follows that,
 even in cases where time is considered as a metrological resource, pure-dephasing  is not the optimal choice. The choice $\theta = \pi/8$ and $\alpha=0$, for example, leads to a globally better rate $q$. This contrasts with the results obtained for short times (\Fref{fig:mapFigs}(a)-(c)), where pure-dephasing dynamics resulted to be always optimal. 

The sub-optimality of pure dephasing dynamics is even more evident when the estimation of the cutoff parameter $\omega_c$ is addressed. Fig.\ref{fig:mapFigs}(e)-(f) already showed that, in this case, there are interaction angles and initial states outperforming pure  dephasing. Fig.\ref{fig:tedopaFig}(d)-(e) show that, when longer times are considered, the choice $\theta=0$ $\alpha=0$ leads to a 33\% larger value of $q(\omega_c,t)$ for $t\approx 1$ and $\omega_S=1$ w.r.t. pure dephasing, whereas the same choice of the interaction and initial state angles leads to a 100\% larger value of $q(\omega_c,t)$ when $\omega_S=5$. Qualitatively similar behaviors are obtained for super- ($s>1$) and sub- ($s<1$) Ohmic spectral densities (not shown).
\section{Conclusions}
In this paper we have investigated whether engineering the interaction 
Hamiltonian may improve the precision of quantum probing. In particular, 
we have considered a qubit probe interacting with a bosonic Ohmic 
environment and have addressed the effects of going beyond pure dephasing on
the precision of estimation of environmental parameters such as the
temperature, or the cutoff frequency of the environment spectral density.
We have analyzed the behavior of the maximal extractable information, 
as quantified by the quantum Fisher information, for different
initial preparations of the probe and system-bath interaction.
\par
Our results provide clear evidence that pure dephasing interaction is not optimal in general, except for very short times. The presence of a transverse interaction 
may indeed improve the maximum attainable precision in several working regimes.  
From a physical point of view, our results show that the exchange of energy 
between the system and the environment plays a major role in determining the 
QFI and this is especially evident in the strong-coupling regime, see e.g. Figure \ref{fig:tedopaFig}}.
\par
Besides the dynamics, we have also analyzed the role of the kinematics of the probe
in determining the precision of the estimation. In particular, we have analyzed
the role of the initial state of the probe and that of its characteristic 
frequency. Our results illustrate the complex interplay among the different features of the probe and provide quantitative guidelines to design optimal detection schemes characterizing bosonic environments at the quantum level.
\section*{Acknowledgements}
We thank Andrea Smirne for useful discussions. MGAP is member of INdAM-GNFM. HPB acknowledges support from the Joint Project ``Quantum Information Processing in Non-Markovian Quantum Complex
Systems'' funded by the Freiburg Institute for Advanced Studies (FRIAS,
University of Freiburg) and the Institute of Advanced Research (IAR,
Nagoya University). DT acknowledges support from the University of Milan through the ``Sviluppo UniMi'' initiative.
\appendix
\section{Generator and dynamical map}\label{appa}
The matrix form of the generator $\mathcal{L}(t)$, defined in \eref{eq:tcl} with coefficients given by \eref{eq:tclcoeff} can be derived by using  \eref{eq:tomatrix} and keeping only terms up to $t^3$. 
\begin{equation}
D^{\mathcal{L}(t)}_{(3)} = \left (
\begin{array}{c|c}
0 & \boldsymbol{0} \\
\hline
\boldsymbol{\mu}_{(3)}(t) & W_{(3)}(t)
\end{array}
 \right),
\end{equation}
with $\boldsymbol{0}$ the three-dimensional zero vector,
\begin{equation}
\boldsymbol{\mu}_{(3)}(t) = \frac{t^3 \zeta(1) \omega_S}{6} \left (\sin (2 \theta),0, -2 \cos^2(\theta) \right )^T,
\end{equation}
and 
\begin{widetext}
\tiny
\begin{equation}
W_{(3)}(t) = \left ( 
\begin{array}{ccc}
  \frac{1}{6} \zeta (2) t^3 \sin ^2(\theta )-\zeta (0)
   t \sin ^2(\theta ) & -\omega_S & \frac{1}{2}
   \zeta (0) t \sin (2 \theta )-\frac{1}{12} t^3
   \left(\zeta (0) \omega_S^2+\zeta (1)
   \omega_S+\zeta (2)\right) \sin (2 \theta ) \\
  \frac{1}{2} t^2 \omega_S (\zeta (0) \cos^2(\theta) )+\omega_S & \frac{1}{6} t^3
   \left(\zeta (0) \omega_S^2 \cos^2 (\theta
   )+ \zeta (2)\right)-\zeta
   (0) t & \frac{1}{8} t^2 (2 \zeta (0) \omega_S+\zeta (1)) \sin (2 \theta ) \\
  \frac{1}{12} t^3 (\zeta (1) \omega_S-\zeta (2)) \sin (2 \theta )+\frac{1}{2} \zeta (0) t
   \sin (2 \theta ) & -\frac{1}{8} \zeta (1) t^2 \sin (2
   \theta ) & \frac{1}{6} t^3 \left(\zeta (0) \omega_S^2+\zeta (2)\right) \cos ^2(\theta )-\zeta (0) t \cos
   ^2(\theta )
\end{array}
\right) 
\end{equation}
\normalfont
\end{widetext}
It is worth noticing that the generator of the translations $\bs{\mu}_{(3)}(t)$ depends on the first moment $\zeta(1)$ of the spectral density. It is possible to show, by direct inspection of higher-order generators $D^{\mathcal{L}(t)}_{(k)}, \ k>3$ that the generators of the translations depend only on the odd-moments  $\zeta(2n+1)$. For spectral densities belonging to the Ohmic family, such odd moments are independent of the temperature.

\section{TEDOPA algorithm}\label{tedopaApp}
To simulate the evolution of the spin-boson model, we resorted to the the recently proposed Thermalized Time Evolving density matrix with orthogonal polynomials (T-TEDOPA) algorithm. In this section we briefly present the T-TEDOPA scheme and refer to \cite{tamascelli19} for
a more detailed presentation of the algorithm. {Clearly enough, other numerical methods such as Hierarchical Equation of Motion (HEOM) \cite{tanimura89,tanimura06}, or the recently proposed Transformation to Auxiliary Oscillators (TSO) \cite{tama20,tamascelli18}, can be applied, as long as high enough accuracy is guaranteed}.

T-TEDOPA is a certifiable and numerically exact method \cite{prior10, woods14, woods2015}
to efficiently treat finite-temperature open quantum system dynamics.
T-TEDOPA first extends the bosonic environment by including negative frequency modes. The initial state of the extended enviroment, governed by the Hamiltonian $H_E^{\text{ext}}= \int_{-\infty}^{+\infty} d\omega \ a_\omega^\dagger a_\omega$, is set to the (pure) vacuum state $\ket{0}_E$ (i.e. $a_\omega \ket{0}_E = 0\  \forall \omega \in \mathbb{R}$). The spectral density $J(\omega)$ is then replaced by a the \emph{thermalized} spectral density 
\begin{equation}
J_\beta(\omega) = \frac{J^{\text{ext}}(\omega)}{2}\llrrq{1+\coth\llrr{\frac{\beta \omega}{2}}}
\end{equation}
with $J^{\text{ext}}(\omega) = \text{sign}(\omega) J(|\omega|)$. Since $J_\beta(\omega)$ is a measure, i.e a positive valued function, on $\mathbb{R}$, it is possible to determine a family of polynomials $p_{\beta,n}(\omega)$ orthogonal w.r.t. the measure $d\mu_\beta = J_\beta(\omega)d\omega$, and define new creation and annihilation operators $c_{n,\beta}^{(\dagger)}$ through a unitary transformation:
\begin{align}
U_{\beta,n}(\omega) &=\sqrt{J_\beta(\omega)} p_{\beta,n}(\omega), \\
c_{\beta,n}^{(\dagger)} &= \int_{-\infty}^{+\infty} d\omega \ U_{\beta,n}(\omega) a_\omega^{(\dagger)} .
\end{align}
As for standard TEDOPA, thanks to the three-term recurrence relation satisfied by the polynomials $p_{\beta,n}(\omega)$, the $H_{SE}(\theta)$ Hamiltonian \eref{eq:completeHam} is mapped into a chain Hamiltonian $H^C(\theta) = H_S+H_E^C+H_I^C(\theta)$ where
\begin{align}
&H_I^C(\theta) = \kappa_{\beta,0} A(\theta)(c_0+c_0^\dagger) \\
&H_E^C = \sum_{n=0}^{+\infty} \omega_{\beta,n} c_n^\dagger c_n + \sum_{n=1}^{+\infty} \kappa_{\beta,n} (c_{n-1}c_n^\dagger +c_{n-1}^\dagger c_n),
\end{align}
with $A(\theta)$ defined as in \eref{eq:interaction}. The transformation therefore maps the environment  into a semi-infinite one-dimensional chain of oscillators with nearest-neighbor interactions and the coefficients  $\omega_{\beta,n}, \ \kappa_{\beta,n}$ are, repsectively, the temperature dependent chain oscillators frequencies and coupling strengths, directly related to the coefficients of the recurrence relation defined by the orthogonal polynomials $p_{\beta,n}(\omega)$. These latter are typically computed by means of stable numerical routines \cite{gautschi94}.
\begin{figure}
\includegraphics[width=1.\columnwidth]{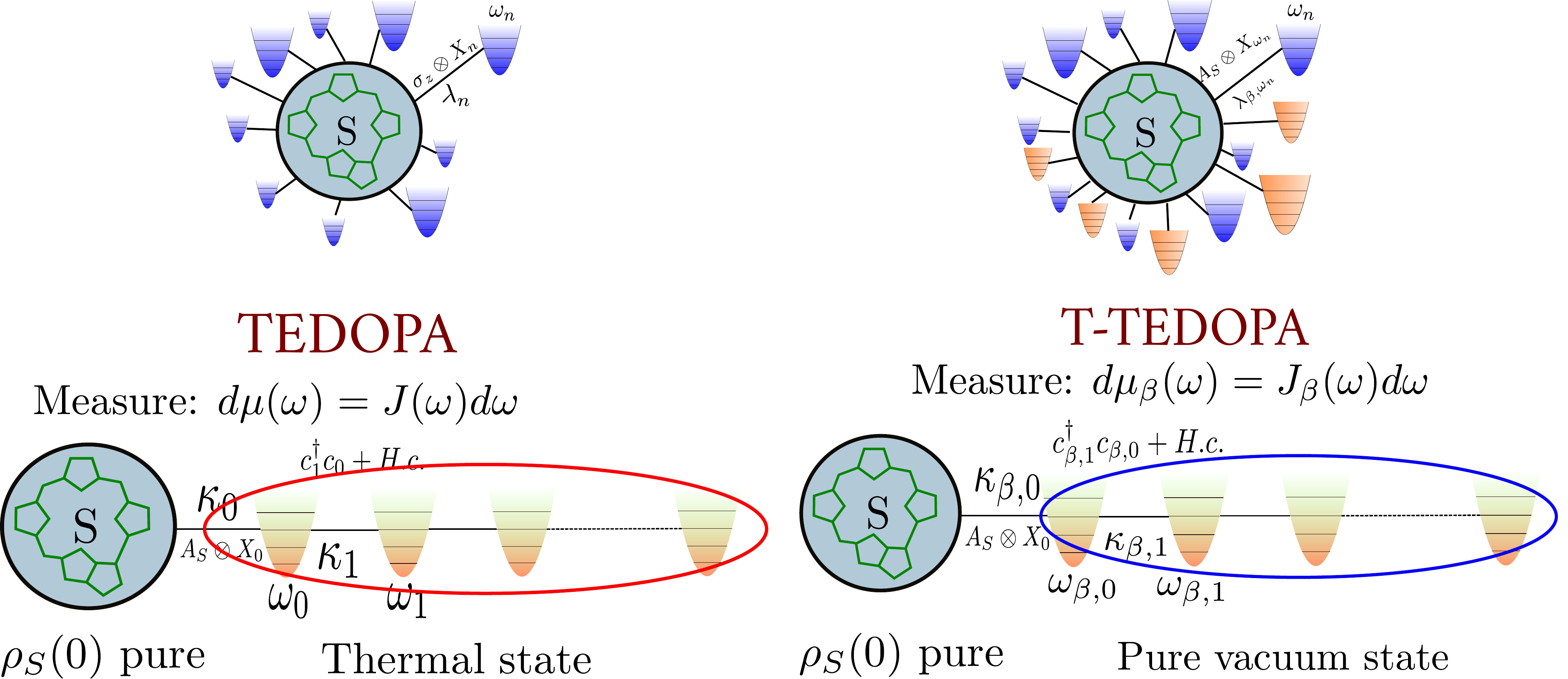}
\caption{\label{fig:tedopa}A pictorial scheme of the TEDOPA and T-TEDOPA transformations.}
\end{figure}
This transformation from the spin-boson model to a one-dimensional geometry is depicted in Fig.~\ref{fig:tedopa}.

In a second step this emerging configuration is treated by Time Evolving Block
Decimation (TEBD) method. 
TEBD  generates a high fidelity approximation of the time evolution of a one-dimensional
system subject to a nearest-neighbor Hamiltonian with polynomially scaling computational resources. TEBD
does so by dynamically restricting the exponentially large Hilbert space
to its most relevant subspace thus rendering the computation feasible \cite{vidal04}.

TEBD is essentially a combination of an MPS description  \cite{schollwoeck11} for a one-dimensional
quantum system and an algorithm that applies two-site gates that are necessary
to implement a Suzuki-Trotter time evolution \cite{suzuki90}. Together with MPS operations such
as the application of measurements this yields a powerful simulation framework.
An extension to mixed states is
possible by introducing a matrix product operator (MPO) to describe the density
matrix, in complete analogy to an MPS describing a state
\cite{zwolak04,schollwoeck11}. Such an extension is indeed not needed in our simulations. As a matter of fact we consider only pure initial states of the system. The environmental initial state is, instead, a thermal state. However, by applying T-TEDOPA, we are able to shift the thermal contributions from the initial state of the chain to temperature-dependent chain coefficients, and initialize the chain in the (pure) vacuum state. This provides us with the possibility of using a pure state (MPS) description of the overall system-environment state, with major computational advantage. We refer to \cite{tamascelli19} for a more detailed comparison between T-TEDOPA and TEODPA.

A last step  is necessary to adjust this configuration further to suit
numerical needs. The number of levels for the environment oscillators can be restricted to a
value $d_{\text{max}}$ to reduce required computational
resources. A suitable value for $d_{\text{max}}$ is related to the
sites average occupation which, in turn, depends on the environment structure and
temperature. In our simulations we set $d_{\text{max}}=12$: this value provides converged results
for all the examples provided. The Hilbert space dynamical reduction performed by TEBD is
determined
to the \emph{bond dimension}. The optimal choice of this parameter depends on the amount of
long range correlations in the system. For all the simulations used in this work, a bond
dimension $\chi=50$ provided converged results.  At last, we observe that the mapping described
above produces a semi-infinite chain that must be truncated in order to enable simulations.  In
order to avoid unphysical back-action on the system
due to finite-size effects, i.e. reflections from the end of the chain, the chain has to be
sufficiently long to completely give the appearance of a ``large'' reservoir. These truncations can
be  rigorously certified by analytical bounds \cite{woods15}. For the examples provided in the
paper, chains of $n=150$ sites are more than enough to see no boundary effect.

As to further optimize our simulations, we augmented our TEDOPA code with a  Reduced-Rank Randomized Singular Value  Decomposition (RRSVD) routine \cite{tama15,kohn18}.  Singular value decomposition is at the heart of the dimensionality reduction TEBD relies on.  RRSVD is a randomized version of the SVD that provides an improved-scaling SVD, with the same accuracy as the standard state of the art  deterministic SVD routines.

%

%
%

\end{document}